\documentclass[twocolumn,letterpaper]{IEEEAerospaceCLS}  % only supports two-column, letterpaper format

\usepackage[]{graphicx}    % We use this package in this document
\usepackage{braket}
\usepackage{amsmath}
\usepackage{amsfonts}
\usepackage{enumitem}
\usepackage{mathtools}
\usepackage{blkarray, bigstrut} 
\usepackage{bbm}   %\usepackage{dsfont}

\newcommand{\ignore}[1]{}  % {} empty inside = %% comment
\newcommand{\ppdd}{p_d}
\newcommand{\binsett}{{\mathbb{{B}}}}   % {{\mathbb{{B}}}}   %  {{\{0,1\}}}
\newcommand{\X}{{\mathcal{X}}}      
\newcommand{\Y}{{\mathcal{Y}}}      

\newcommand{\costQ}{{\mathbf{\Gamma}}}
\newcommand{\kalmanQ}{{Q_{{\text{proc}}}}}

\newcommand{\measspaceY}{[0,100]}
\newcommand{\lengthmeasspaceY}{100} % length of interval above
\newcommand{\fov}{|\text{FoV}|}

\newcommand{\meas}{y}
\newcommand{\measScanK}{{{\meas}_k}}

\newcommand{\exampleAssocMatrix}{{\mathbf{{S}}}}   %{{\mathbf{{S}}_{\scriptscriptstyle \rm{feas}}}}   %{{\mathbf{{S'}}}}

\newcommand{\WrFromKrooks}{W_r}   %{{{\mathbf{W}}_r}}
\newcommand{\WcFromKrooks}{W_c}   %{{{\mathbf{W}}_c}}
\newcommand{\thetaRKrooks}{\theta_r}
\newcommand{\thetaCKrooks}{\theta_c}
\newcommand{\WrFromMTDA}{W'_r}   %{{{\mathbf{W}'}_r}}  
\newcommand{\WcFromMTDA}{W'_c}   %{{{\mathbf{W}'}_c}}
\newcommand{\thetaRMTDA}{\theta'_r}
\newcommand{\thetaCMTDA}{\theta'_c}

\newcommand{\qkrooks}{{q_{\scriptscriptstyle \rm{kR}}}}
\newcommand{\qqkrooks}{{Q_{\scriptscriptstyle \rm{kR}}}}
\newcommand{\qmtda}{{q_{\scriptscriptstyle \rm{MTDA}}}}
\newcommand{\qqmtda}{{Q_{\scriptscriptstyle \rm{MTDA}}}}

\newcommand{\qmtdaVec}{B}   %{{B_{\scriptscriptstyle \rm{MTDA}}}}
\newcommand{\qmtdaVecEntry}{{B}}

\begin{document}
\title{Implementation of a Multiple Target Tracking Filter on an Adiabatic Quantum Computer}

\author{%
Timothy M. McCormick, Bryan R. Osborn, R. Blair Angle, Roy L. Streit \\ 
Metron, Inc.\\
1818 Library St., Suite 600\\
Reston, VA 20190\\
\{McCormickT, Osborn, Angle, Streit\}@metsci.com
%%%% IMPORTANT: Use the correct copyright information--IEEE, Crown, or U.S. government. 
%%%%%
\thanks{\footnotesize 978-1-6654-3760-8/22/$\$31.00$ \copyright2022 IEEE}              % This creates the copyright info that is the correct 2021 data.
}

\maketitle

\thispagestyle{plain}
\pagestyle{plain}

\maketitle

\thispagestyle{plain}
\pagestyle{plain}

\begin{abstract}
Recent  work at Fraunhofer FKIE shows  that  Morefield's method for multiple target data association 
\cite{mroefield} can in  theory  be solved on an adiabatic quantum computer. The present paper validates the theory and examines the significant limitations of  currently available adiabatic quantum computers  for solving the data association problem.  The limitations of such architectures are both theoretical and practical in nature, and both are discussed. The data association problem is formulated as a quadratic unconstrained binary optimization (QUBO) problem; consequently, much of the  discussion is  relevant to other applications which are, or can be, posed as QUBO problems. 
\end{abstract}

\tableofcontents

%%%%%%%%%%%%%%%%%%%%%%%%%%%%%%%%%%%%%%
\section{Introduction}
%%%%%%%%%%%%%%%%%%%%%%%%%%%%%%%%%%%%%%
The promise of quantum computing (QC) is widely touted in diverse fields, and it attracts significant levels of attention from many quarters.  This paper is about data association problems.  Various forms of this problem  lie at the very heart of many target tracking problems.  One form of the tracking problem is called herein the multi-target data association (MTDA) problem, and it  was recently shown theoretically \cite{mtda} to be solvable by adiabatic quantum computers.  Adiabatic quantum computation (AQC) \cite{qa1,qa2,qa3} is based on quite a different idea from traditional gate-based models for quantum computation \cite{deutsch1}.  In AQC, the computation executes by evolving an initial Hamiltonian with an easy to prepare ground state to a final Hamiltonian whose ground state encodes the solution to the computational problem of interest.  

The insight provided by  \cite{mtda} is to cast the quadratic unconstrained binary optimization (QUBO) problem for the MTDA as an Ising model that can be represented in terms of the qubits on an adiabatic quantum computer.  The present paper presents the results of implementing that Ising model on a class of AQCs called  D-Wave machines \cite{martinisDwave} and solving for the solution to the underlying QUBO.  Our work validated the theory in Ref. \cite{mtda}, but it also uncovered many issues that govern and/or limit  performance on AQCs.  These issues are candidly discussed in this paper.  While many  will undoubtedly be overcome  as the technology advances,  others are physics-based and  pose more serious challenges.      

Section \ref{dataassoc} reviews the MTDA problem.  Section \ref{adaibaticMTDA} defines the time-dependent Hamiltonian needed to solve the QUBO for  MTDA on a D-Wave machine.  It also discusses issues that limit performance and scalability.   Section \ref{results}  presents the results of using D-wave to solve what is called the $k$-rooks problem.  This  is the MTDA problem with no false alarms and no missed detections.   Section \ref{MTDAresults} presents results for  the general MTDA problem.  This section is the main application interest in the paper. It builds on the $k$-rooks discussion. The results show that the MTDA problem is, in many ways, an ideal problem to begin to understand QC more generally.    Section \ref{numerical} presents results of a numerical study of the choice of anneal time for a small MTDA system that can be solved numerically.   Section \ref{issues} gives examples showing that the minimum energy shot (for fixed length runs) follows Gumbel's extreme statistics distribution.    Section \ref{hybrid} discusses the need to solve hard problems in practice will probably require a combination of quantum and digital computers.   Section \ref{conclsion} gives concluding remarks.

%%%%%%\newpage

%%%%%%%%%%%%%%%%%%%%%%%%%%%%%%%%%%%%%%%%%%%
\section{Data Association and Tracking}\label{dataassoc}
%%%%%%%%%%%%%%%%%%%%%%%%%%%%%%%%%%%%%%%%%%%

%%% WE CAN REMOVE THESE NEWCOMMANDS LATER. TOO LAZY TO DEAL WITH RIGHT NOW.
\newcommand{\kalmanPzero}{P_{0 \givennn 0}}
\newcommand{\kalmanPkPred}{P_{k \givennn k-1}}
\newcommand{\kalmanPkMinusOne}{P_{k-1 \givennn k-1}}
\newcommand{\given}{\,|\,}
\newcommand{\givennn}{|} 
\newcommand{\qijVectorTerm}{y_k^j - HF \hat{x}^i_{{{k-1 \givennn k-1}}}}
\newcommand{\SHORTqijVectorTerm}{{d_{ijk}}}
\newcommand{\kapp}{{n_d}}   %  \kappa   {{n_d}}

The tracking problems of interest here are those in which a sensor produces multiple measurements at a succession of scans. When false alarms (i.e., clutter) are present, and when multiple targets are present but may or may not be detected, the core of the tracking problem becomes that of deciding which measurement should be assigned to which target or, if not to a target, then to clutter. This is the multiple target data association (MTDA)  problem. MTDA is clearly  a combinatorial problem.   

Two kinds of tracking paradigms are commonly used to solve the problem.  One is called probabilistic data association (PDA)  filtering, and it tries to avoid making hard  assignments by estimating the probabilities of assignments and using them to approximate a tracking filter.  The approach  is sometimes called ``soft'' assignment. 

The other method explicitly seeks to find the best assignments, and then builds the tracking filter conditioned on the assignments.  This approach is often called ``hard'' assignment.  The approach leads to  combinatorial optimization problems.  One  such problem takes the form of a binary integer linear programming (ILP) problem.  As seen in the Appendix, ILPs can be solved by adiabatic Quantum Computation (QC). The recognition that  MTDA problems can be solved by QC  was first reported in \cite{mtda}. 

The JPDA filter is discussed first because it sets the stage for stating the hard association binary ILP problem that is the real focus of this paper. The discussion here is limited to one scan.

\subsection{Joint Probabilistic Data Association} 
The joint probabilistic data association (JPDA) filter \cite{barshalomjpda} 
is a Bayesian filter for tracking a known
number of targets. The   sensor produces point measurements that are modeled as the superposition of point target measurements and  points from an independent clutter process in 
a common measurement space $\Y$.
%A target that generates a  measurement is said to be detected by the sensor.
Superposition models the fact that measurements are unlabeled, i.e., 
whether a given measurement is clutter-induced or generated 
by one of the $N \geq 1$ targets is unknown.
 Targets are
assumed independent of each other (each target $i$ has its own state space $\X^i$) 
and causally independent of the measurement process. 
Each target generates at most one
measurement in any given scan.
 Measurements are assigned to at most one target, and the measurement
likelihood function depends only on the target to which it is
assigned. Measurements not assigned to a target are assigned
to the clutter process. For any given scan, the scan measurement set may be empty.

On a classical computer, JPDA is not practical in situations where the number of targets $N$ and/or the number of 
measurements $M$ in a given scan are large. The bottleneck is that the exact JPDA requires computing all the feasible  association probabilities, and enumerating them all is, in general, an NP-hard problem \cite{poore} \cite{uhlmann} that is closely related to calculating the matrix permanent \cite{valiant}. For further discussion of the computational complexity of the JPDA filter and its implementation see, e.g., \cite{barshalomjpda},  \cite{ybsANDli},   \cite{metronacbook}.

The specific tracking example used in this paper is now described. All units of length are in meters, and all time units are in seconds. 
%{\color{blue}(Say that regular scans is not a rqmt of JPDA?)}
Scans occur at one second intervals, beginning at time $t_1 = 1$; i.e., $\Delta t = 1$. 
Thus, the $k$-th scan occurs at time $t_k = k$ 
(the scan time is defined to be the end time of the scan interval). 
For scan $k$ and target $i$, $i=1,\ldots,N$, the corresponding (ground truth) target state is denoted $x_k^i \in \X^i \equiv \X \subset {\mathbb{R}}^2$.
The state vector comprises a spatial and a velocity component and is of the form $(p,\dot{p})$.
At the reference time, $t_0 = 0$, no measurements are available and
target $i$ is assumed to have prior PDF $\mu_0(x_0^i)  = {\mathcal{N}}\!\!\left(x_0^i ; \hat{x}_{{0 \givennn 0}}^i , \kalmanPzero^i \right)$, where
 ${\mathcal{N}}(x ; \mu , \Sigma)$ represents the PDF of a multivariate Gaussian with mean vector $\mu$ and covariance matrix $\Sigma$ evaluated at $x$. 
%%Due to the assumption of target independence, the joint prior PDF for all target states is the product over $i$ of the marginals $\mu_0(x_0^i)$.

At scan $k$, the linear-Gaussian motion model for target $i$ is given by
\begin{equation}     \label{eq:jpdaMotUpdate}    %%%\label{eq:jpdaTargetPDF}
    p(x_k^i \given x_{k-1}^i)   =  {\mathcal{N}}\!\left(x_k^i ; F x_{k-1}^i , \kalmanQ \right) ,
\end{equation}
where the process (motion) matrix 
$     %\begin{equation}
F = 
\begin{pmatrix}
1 \,\, & \Delta t \\
 0 \,\,   &  1
\end{pmatrix}  \!, 
$         %\end{equation}
and 
$     %\begin{equation}
\kalmanQ  = 
\sigma^2_p\begin{pmatrix}
 \frac{\Delta t^3}{3} &  \frac{\Delta t^2}{2}\\
  \frac{\Delta t^2}{2} &  \Delta t
\end{pmatrix}  %%%\!, 
$         %\end{equation}
%%%and $\kalmanQ \in {\mathbb{{R}}}^{{2 \times 2}}$ 
is the (specified) process noise covariance matrix.  We take $\sigma_p^2 = 1$.  These matrices are independent of time index $k$.

A target that generates a  measurement is said to be detected by the sensor.  
The measurement space in our example is   $\Y = \measspaceY \subset {\mathbb{R}}$. $\Y$ is a bounded 1-D space, and  we define the clutter (false alarm) PDF to be uniform over $\Y$.   We assume a constant probability of detection over the entire state space $\X$ and over all scans for all targets;
i.e., $P\!d_k^{\,i}(x_k^i) \equiv \ppdd = 0.95$ for all $i,k$.      
Given a detection $y$ generated by target $i$, the measurement likelihood is given by
\begin{equation} \label{eq:jpdaLF}
p(y \given x_k^i) = {\mathcal{N}}\!\left(y ; H x_k^i , \sigma_M^2 \right) , 
\end{equation}
where the measurement matrix $H = 
\begin{pmatrix}
 1 \,\,   &  0
\end{pmatrix}$ 
extracts the spatial component of $x_k^i$,
and $\sigma_M^2 = 0.1 \text{ m}^2$ is the (specified) measurement variance.

Targets are assumed independent, so the joint prior PDF for all target states, that is, the joint target state, at
reference time $t_0=0$ is the product of the prior PDFs  $\mu_0(x_0^i)$:
\begin{equation}
p_0(x_0^1,\ldots,x_0^N) = \prod_{i=1}^N \mu_0(x_0^i) = \prod_{i=1}^N {\mathcal{N}}\!\left(x_0^i ; \hat{x}^i_{{{0 \givennn 0}}} ,\,
   P_{0 \givennn 0}^i \right).
\end{equation}
As will be seen, the JPDA filter imposes this factored form at each step $k$ of the  recursion.  This is an approximation, and it  ensures the posterior distribution has the exact same mathematical form as the prior distribution.  In other words, the approximation  closes the Bayesian  recursion. 

The predicted state for target $i$ at scan $k \geq 1$ is given by
\begin{equation}
    \mu_k^{i-}(x_k^i) ={\mathcal{N}}\!\!\left(x_k^i ; F \hat{x}^i_{{{k-1 \givennn k-1}}} ,\kalmanPkPred^i \right)   ,
\end{equation} 
where 
\begin{equation}
    \kalmanPkPred^i = F  \kalmanPkMinusOne^i F^T + \kalmanQ  \,. 
\end{equation}
 Note that if the initial state covariances $\kalmanPzero^i \equiv \kalmanPzero$ for all $i$, then $\kalmanPkPred^i \equiv \kalmanPkPred$ does not depend on target index $i$.

Let $y_k = \{y_k^1, \ldots, y_k^M\}$ be the measurement set at scan $k$. If it is known which of these measurements, if any, are generated by which target, then the Bayesian information  update would be very simple---it would comprise $N$ classical Kalman filters that process the  measurement generated by the target.  The MTDA assignment problem arises because these assignments are unknown.

The exact JPDA information update is a weighted sum over all the feasible assignments. The weights are the assignment probabilities. The sum is very large in general, and too cumbersome to write explicitly here.  An explicit expression for the exact posterior JPDA assignment probabilities is given in many references, e.g.,    \cite[Eqn. (3.15)]{metronacbook}.  

To close the Bayesian recursion at scan $k \geq 1$, the JPDA filter  approximates the large sum that is the {\textit{joint}} posterior 
target state distribution with  a product of the individual object marginal PDFs
 (known in some circles as a \emph{mean field approximation}).
 Under the current assumptions, in conjunction with the clutter 
 model outlined in the next paragraph, the marginal posterior PDF 
 for target $i$ turns out to be a 
 Gaussian mixture,
 which is then approximated by a single Gaussian with the same mean 
 $\hat{x}^i_{{{k \givennn k}}}$ and 
 covariance $P_{k \givennn k}^i$
 as the mixture; see \cite{metronacbook} for details.
 In particular, 
 given the measurement set $y_k$,
 the joint posterior PDF at scan $k$ is assumed to be of the form
\begin{equation}
p_k(x_k^1,\ldots,x_k^N | \meas_k) = \prod_{i=1}^N {\mathcal{N}}\!\left(x_k^i ; \hat{x}^i_{{{k \givennn k}}} ,\, P_{k \givennn k}^i \right)    ,
\end{equation}
and the Bayesian recursion is closed.

The MTDA paper \cite{mtda}  incorporates all of these standard JPDA assumptions.
In contrast with the MTDA paper, however, here we generalize the clutter model
to a homogeneous Poisson point process (PPP) with mean $\lambda$. 
More specifically, at each scan the number, $n_c$, of clutter points is generated according to a Poisson probability mass function (PMF) with mean $\lambda$. If $n_c>0$, the clutter points are then uniformly and independently distributed over the volume of the entire field of view, denoted $\fov$. As the measurement space $\Y$ is one-dimensional in the current scenario, 
its ``volume" is actually a length and is given by $\fov = \text{len}(\Y) = \text{len}(\measspaceY) = \lengthmeasspaceY$.
Setting $\lambda = 1$ 
reduces this generalized model to that of the MTDA paper,
and this value of $\lambda$ is assumed in the remainder of this paper.

\subsection{Hard Data Association}  Morefield's approach to MTDA  \cite{mroefield}   seeks to find the {\textit{best}} association of measurements to targets or to clutter. It replaces probabilistic data association with a combinatorial optimization problem. This problem becomes a binary integer linear programming problem (ILP) by invoking Drummond's ``at most one measurement per target per scan rule.''  This ILP is equivalent to a QUBO, which in turn is solved by the adiabatic QC machine called D-Wave.

Define $\binsett \equiv \{0,1\}$. Given $N \geq 1$ targets and $M \geq 0$ measurements, we now define the binary association matrix 
  ${\mathbf{{S}}} = \left(S_{ij}\right) \in {\binsett}^{(N+1)\times(M+1)}$. 
  For convenience, the rows of $\mathbf{{S}}$ are indexed by $i=0,1,\ldots,N$ and 
  the columns are indexed by $j=0,1,\ldots,M$. A nonzero entry $S_{ij}$ for $i,j>0$ indicates an association between target $i$ and measurement $j$.
  A nonzero entry $S_{ij}$ for $i=0, j>0$ indicates that measurement $j$ is a clutter measurement. 
  A nonzero entry $S_{ij}$ for $i>0, j=0$ indicates that target $i$ is not detected. The entry $S_{00}$ 
  is there simply to complete the matrix; 
  it is of no consequence and can be discarded from the final solution.
  A \emph{feasible} association matrix is one which satisfies the measurement assignment constraints dictated by JPDA, 
  as outlined in the beginning of this section.

An example of a feasible association matrix 
  $\exampleAssocMatrix = \left(S_{ij}\right)$ % ${\mathbf{{S}}}_{{\text{ex}}}$ 
  for $N=3$ and $M=4$ is given 
  in Eqn. (\ref{feasibleAssocMatExample}). In this example, targets 1 and 3 generate measurements 1 and 3, respectively, while target 2 
  is undetected. Measurements 2 and 4 are false alarms (clutter).  
  \begin{align}   
  \exampleAssocMatrix  &  \,=\,
  \begin{blockarray}{*{5}{c} l}
    \begin{block}{*{5}{>{$\footnotesize}c<{$}} l}
      no det & meas 1 & meas 2 & meas 3 & meas 4 & \\
    \end{block}
    \begin{block}{(*{5}{c})>{$\footnotesize}l<{$}}
      S_{00} & S_{01} & S_{02} & S_{03} & S_{04} \bigstrut[t]& \,FA \\
      S_{10} & S_{11} & S_{12} & S_{13} & S_{14}  & tgt 1 \\
      S_{20} & S_{21} & S_{22} & S_{23} & S_{24}  & tgt 2 \\
      S_{30} & S_{31} & S_{32} & S_{33} & S_{34}  & tgt 3 \\
    \end{block}
  \end{blockarray}     \nonumber   \\
      &  \,=\, \begin{blockarray}{*{5}{c} l}
    \begin{block}{*{5}{>{$\footnotesize}c<{$}} l}
        \quad\quad\quad\,  & \quad\quad\quad & \quad\quad\quad\, & \quad\quad\quad & \quad\quad\quad\, &  \, \\
    \end{block}
    \begin{block}{(*{5}{c})>{$\footnotesize}l<{$}}
      * & 0 & 1 & 0 & 1  \bigstrut[t]&   \\
      0 & 1 & 0 & 0 & 0  &   \\
       1 & 0 & 0 & 0 & 0  &   \\
      0 & 0 & 0 & 1 & 0  &   \\
    \end{block}  
  \end{blockarray}       \label{feasibleAssocMatExample}
    %  &= \begin{pmatrix}  * & 0 & 1 & 0 & 1 \\  0 & 1 & 0 & 0 & 0 \\  1 & 0 & 0 & 0 & 0 \\  0 & 0 & 0 & 1 & 0 
    %  \end{pmatrix}     \label{feasibleAssocMatExample}
   \end{align}    
Let $\kapp \equiv \sum_{i,j>0} S_{ij} \leq \text{min}\{M,N\}$ 
represent the number of target detections associated with a feasible association matrix 
${\mathbf{{S}}} = \left(S_{ij}\right)^{0 \leq i \leq N}_{0 \leq j \leq M}$. 
Restricting Eqn. (3.15) of \cite{metronacbook}
to the conditions of this paper gives the following posterior association likelihood:
  \begin{align} 
  L(\mathbf{{S}} | \measScanK) &= \left(\frac{\lambda}{\fov}\right)^{M-\kapp} \left(1-\ppdd\right)^{N-\kapp}\ppdd^\kapp    \label{ACthreefifteen} \\
  &\,\,  \times  \prod_{\substack{    {S_{ij}=1} \\ i,j>0}}  
    \int_{\X^i} \mu_k^{i-}(x_k^i)p_k^i(\meas_k^j | x_k^i) \text{d}x_k^i    \,,  \nonumber
  \end{align}
  where the product is taken to be $1$ if $n_d = 0$.

%%%Under the assumptions listed above, after removal of unnecessary constants and taking the negative log likelihood, the more general posterior of Eqn. (3.15)
Taking the negative logarithm reduces the likelihood in Eqn. (\ref{ACthreefifteen})
to the cost matrix 
$\costQ = (\Gamma_{ij})$ (called $Q_{ij}$ in Eqn. (40) of the MTDA paper and renamed here to avoid confusion with the cost matrix introduced later), with the exception
of an additional $\lambda$ term. The modified cost matrix is given by
  \begin{equation}  \label{modifiedeqnforty}
    \Gamma_{ij} = 
    \begin{cases}
      -\text{log } \ppdd + \gamma_{ij}, & \text{if}\ i>0, j>0 \\
      \text{log} \left(\frac{\fov}{\lambda}\right), & \text{if}\ i=0, j>0 \\
      %%%%%%\text{log} \fov -\text{log} \lambda, & \text{if}\ i=0 \\
      -\text{log } (1-\ppdd), & \text{if}\ j=0   \,,
    \end{cases}
  \end{equation}
  where
 \begin{equation}
   \gamma_{ij} = \frac{1}{2} \SHORTqijVectorTerm^T
     \left({S}_k^{i}\right)^{-1} \SHORTqijVectorTerm + \frac{1}{2} \text{log } |2 \pi {{S}}_k^{i}|  \,,  
 \end{equation}
 and the Kalman innovation residual and covariance 
 for target $i$ are given by, respectively,
  \begin{equation}
      \SHORTqijVectorTerm = \qijVectorTerm   ,
 \end{equation}
 \begin{equation}
     S_k^{i} = H \kalmanPkPred^i H^T +  \sigma_M^2 .  %R .
 \end{equation}   
(The innovation covariance should not be confused with the emboldened 
association matrix $\mathbf{{S}}$.)
Note that for $0<\ppdd<1$, the third equality in Eqn. (\ref{modifiedeqnforty})
 enforces the constraint $S_{00}=-1$. This forces a choice between two
 solutions ${\exampleAssocMatrix}_1$ and ${\exampleAssocMatrix}_2$ that have 
 different values for $S_{00}$, but are otherwise equal. 
 (If, instead, we were to set $\Gamma_{00}=0$ in Eqn. (\ref{modifiedeqnforty}), then ${\exampleAssocMatrix}_1$ and 
 ${\exampleAssocMatrix}_2$ would have equal costs, which results in ``redundant" solutions.)

 Define the \emph{vectorization} $\text{vec}(A)$ of an $r \times s$ matrix $A$ 
  to be the $rs \times 1$ column vector obtained by concatenating
  the columns of $A$.
  As $x^2 = x$ for $x \in \binsett$, 
  the linear term of the corresponding QUBO 
  (see the Appendix, Eqn.  \eqref{matrixalg})
  is the main diagonal of matrix $Q$ in Eqn. (\ref{qubo}), 
  and it is equal to 
  $\gamma^T s$, where $\gamma = \text{vec}(\costQ) \in {\mathbb{{R}}}^{{(N+1)(M+1)}}$, and $s = \text{vec}({\mathbf{S}})  \in {\binsett}^{(N+1)(M+1)}$.
  %%%%\emph{RBA-- Say something about discarding the $(0,0)$ term when vectorizing. Or simply ignore this term in the solution.}

Using the method outlined in the Appendix, the ILP for MTDA transforms into a QUBO.  This QUBO is the starting point for QC.

%%%%%%%%%%%%%%%%%%%%%%%%%%%%%%%%%%%%%%%%%%%%%
\section{Adiabatic solution of  MTDA}\label{adaibaticMTDA}
%%%%%%%%%%%%%%%%%%%%%%%%%%%%%%%%%%%%%%%%%%%%%

\subsection{Qubits and Assignments}

Following Refs. \cite{mtda,fraunhofer2}, we may write each entry $S_{ij}$ of Eqn. (\ref{feasibleAssocMatExample}) as the state of a two-level system $\ket{\phi_{ij}} \in \{\ket{0},\ket{1} \}$. $\ket{\phi_{ij}}$ can be thought of as one of the basis states of two-dimensional complex Hilbert given by
\begin{equation}
    \label{spinHalfBasis}
    \ket{0} = \begin{pmatrix}
           1 \\
           0 
         \end{pmatrix},\ 
     \ket{1} = \begin{pmatrix}
       0 \\
       1 
     \end{pmatrix},
\end{equation}
which are eigenvectors (with eigenvalues $1$  and $-1$ respectively) of the Pauli matrix $\sigma_3$, which has the following representation in this basis:
\begin{equation}
    \label{spinHalfBasisSigz}
    \sigma_3 = \begin{pmatrix}
           1 & 0\\
           0 & -1 
         \end{pmatrix}.
\end{equation}

The full quantum state for the system in Eqn. (\ref{feasibleAssocMatExample}) can then be written as the Kronecker product
\begin{equation}
    \label{prodState}
    \ket{\Phi_j} = \ket{\phi_{00}} \otimes \ldots \otimes \ket{\phi_{MN}},
    %\ket{\Phi_j} = \ket{\phi_{11}} \otimes \ldots \otimes \ket{\phi_{MN}},
\end{equation}
where $j$ is the integer corresponding to the binary string of association matrix entries:
\begin{equation}
    \label{binaryString}
    j \longleftrightarrow (S_{00}, S_{01}, \ldots, S_{MN}).
    %j \longleftrightarrow (S_{11}, S_{12}, \ldots, S_{MN}).
\end{equation}
The dimension of this product space is equal to $D = 2^{(M+1)(N+1)}$
and an arbitrary state can be written as a linear combination
\begin{equation}
    \ket{\psi(t)} = \sum_{j = 0}^{D-1} c_{j}(t) \ket{\Phi_j},
\end{equation}
which is in general a function of time $t$.

\subsection{Quantum Adiabatics}
The dynamics of a quantum system are governed by the Hamiltonian operator $H(t)$, which generally depends on time. The evolution of the system is given by 
\begin{equation}
    \label{timeevham}
    H(t)\ket{\psi(t)} = i\hbar \frac{d}{dt}\ket{\psi(t)},
\end{equation}
for a quantum state $\ket{\psi(t)}$.
We introduce the instantaneous eigenstates $\ket{n(t)}$ by solving
\begin{equation}
    \label{timeindscheq}
    H(t)\ket{n(t)} = E_{n}(t)\ket{n(t)},
\end{equation}
at the instantaneous time $t$.  The energy of   $\ket{n(t)}$ is the eigenvalue $E_n(t)$.    

The Hamiltonian in Eqn. (\ref{timeevham}) is written as $H(st_f)$, where $t_f$ is the final evolution time and $s \equiv t/t_f \in [0,1]$ is dimensionless.  In adiabatic quantum computing, we  evolve our system from an initial Hamiltonian $H_B$ that is relatively easy to prepare on the device to the final Hamiltonian $H_P$ which characterizes the problem of interest.  At any instant in the scaled time $s  \in [0,1]$, we write the Hamiltonian of interest as the convex combination
\begin{equation}
    \label{adiabham}
    H(s t_f) = (1-s)H_B + s \ H_P.
\end{equation}
If the system begins in an instantaneous eigenstate $\ket{n(t_0)}$ and evolves at a
sufficiently slow rate, then the system will remain in the state with the same eigenindex, up to a phase
$\ket{\psi(t)} = e^{-i\theta(t)}\ket{n(t)}$; that is, it evolves adiabatically. The
evolution of the system with Hamiltonian (\ref{adiabham}) is adiabatic if   the
``energy gap'' $|E_n(s) - E_m(s)|$ between any two instantaneous eigenstates (with labels $n$ and $m$) is sufficiently large  \cite{amin_adiab}:  
\begin{equation}
    \label{adiabThm}
    \frac{1}{t_f} \max_{s\in [0,1]} \frac{|\bra{n(s)} \partial_s H(s)  \ket{m(s)}|}{|E_n(s) - E_m(s)|} \ll 1, \forall m \neq n.
\end{equation}
The arguments that lead to Eqn. (\ref{adiabThm}) are approximate, in the sense that they do not yield strict inequalities.  For more rigorous derivations, we refer the reader to  \cite{adiab_rmp}.

\subsection{QC Solutions Are Probability Mass Functions}  %  (PMFs)}
Due to the exact mapping between quadratic unconstrained binary optimization (QUBO) problems and Ising models (see the Appendix), adiabatic quantum computers can solve QUBOs by casting them as Ising magnet Hamiltonians.  
The binary variables $s_i \in \{ -1,1 \}$ are represented as a set of logical qubits described by Pauli spin operators $\sigma^i_{1},\ \sigma^i_{2},\ \sigma^i_{3}$. (See \cite{mikeandike} for definitions and details.) On adiabatic quantum annealers, the default initial Hamiltonian $H_B$ for an $N_s$ variable QUBO is  \cite{adaibatic}:
\begin{equation}
    \label{dwaveinit}
    H_B = -\sum_{i=1}^{N_s} \sigma^i_{1}.
\end{equation}
The ground state of Eqn. (\ref{dwaveinit}) corresponds physically to a non-degenerate state in which all $N_s$ spins are polarized in the $x$-direction.  This corresponds to an even superposition of the $\sigma_3$ eigenstates, or equivalently, a uniform wavefunction in the computational basis (see Eqn. \ref{prodState}). The system is then adiabatically evolved to the system with Hamiltonian $H_P$, that is, to a system described by the Ising Hamiltonian  
\begin{equation}
    \label{isingHam}
    H_P = \sum_{i,j = 1}^{N_s} Q_{ij}\sigma^i_{3}\sigma^j_{3} + \sum_{i=1}^{N_s} q_i \sigma^i_{3}.
\end{equation}
Hence, at intermediate values of $s$, the Hamiltonian in Eqn. (\ref{adiabham}) is that of a transverse field Ising model \cite{tfim1,tfim2}.
The ground state of Eqn. (\ref{isingHam}) corresponds to the desired solution of the associated QUBO described by $Q_{ij}$ and $q_i$.

\subsection{Quantum Measurements}
Probability is at the heart of quantum mechanics and upon annealing from a system described by Eqn. (\ref{dwaveinit}) to one described by Eqn. (\ref{isingHam}) we obtain a PMF over set of all possible quantum states.  The physics allows one and only one sample of this PMF to be obtained because the act of making a  measurement irretrievably destroys the  phase of the system. The sample is called a projective measurement. The measurement used here is a projection onto the standard  canonical basis \cite{mikeandike}.  (Other kinds of projective  measurements are sometimes useful.) The standard basis uniquely identifies the MTDA  assignments.  In the D-Wave literature, a measurement is also  called  simply  a ``shot.''  

Consequently, to obtain multiple shots of the system PMF,  it is necessary to reinitialize it and perform the full QC annealing process again. When we submit a ``run'' to a D-Wave device, we specify some number of shots $n_s$ (currently limited to $n_s \le 10^4$ per run). To be clear,  each shot represents one anneal from $s=0$ to $s=1$ that  evolves the system Hamiltonian from Eqn. (\ref{dwaveinit}) to Eqn. (\ref{isingHam}).  Since the eigenvalues of the Hamiltonian are energies, it is natural to represent the set of shots by a so-called energy density of annealed states defined by
\begin{equation}
    \label{dos}
    g(E) = \frac{1}{n_s}\sum_{i=1}^{n_s} \delta (E - E_i),
\end{equation}
where $E_i$ is the energy of the $i$-th shot.
As $\delta(0) = 1$ and $\delta(x) = 0$ otherwise, $g(E)$ is depicted in this paper as a normalized histogram, or PMF, over the set of energies $\{E_i\}$.  

\subsection{QUBO Solutions and Extremal Statistics}

As the number of Ising sites $N_s$ grows, the dimension of the Hilbert space spanned by the Ising Hamiltonian grows as $2^{N_s}$, making the classical computation of an exact ground state and its energy $E_0$ challenging.  For even modest values of $N_s$, validating that the quantum annealed solution equals  $E_0$ is impossible in  practice.  As we discuss in the sections that follow, quantum annealing provides a rich landscape of parameters to improve the probability of obtaining the optimal solution with energy $E_0$. Two important parameters are the number of shots $n_s$ and the system annealing time $t_f$. 

We obtain the PMF $g(E)$ over a subset of the eigenstates of a given final Ising model corresponding to the QUBO that we are interested in, however the precise quantitative details of $g(E)$ are not relevant.  Instead, we are interested in the lowest energy state  that is returned by the annealer; thus, we define the minimum energy estimator
\begin{align}
    \hat E_0 = \min \{E_1,\ldots,E_{N_s}\}.
\end{align}
With some probability we obtain the true ground state of the final Ising model with energy we denote $E_0$.  For small systems, it is  possible to verify through exact calculation on a classical computer whether or not $\hat E_{0} = E_0$ by explicitly diagonalizing the final Hamiltonian $H_P$ in Eqn. (\ref{isingHam}). In general, states that are close in energy may not be close together in the $2^{N_s}$ dimensional state space of the Hamiltonian's eigenvectors.

Other approaches to solving Ising problems, such as simulated annealing \cite{simAnneal} using Metropolis MCMC \cite{mcmc} and genetic algorithms, are similarly unable to prove the optimality of their result.  Some recent work has shown that Ising models of a certain type can be transformed into a pair of combinatorial optimization (MAX-SAT) and non-smooth convex optimization problems \cite{isingMaxsat}, however that method is constrained to a specific set of Hamiltonians that does not include our problem.  If we did have a method of efficiently solving for the ground state, then the quantum annealing approach would not be necessary.

\subsection{D-Wave Architecture and Scalability}

\begin{figure}[t]
    \centering
	\includegraphics
	[width=0.45\textwidth]{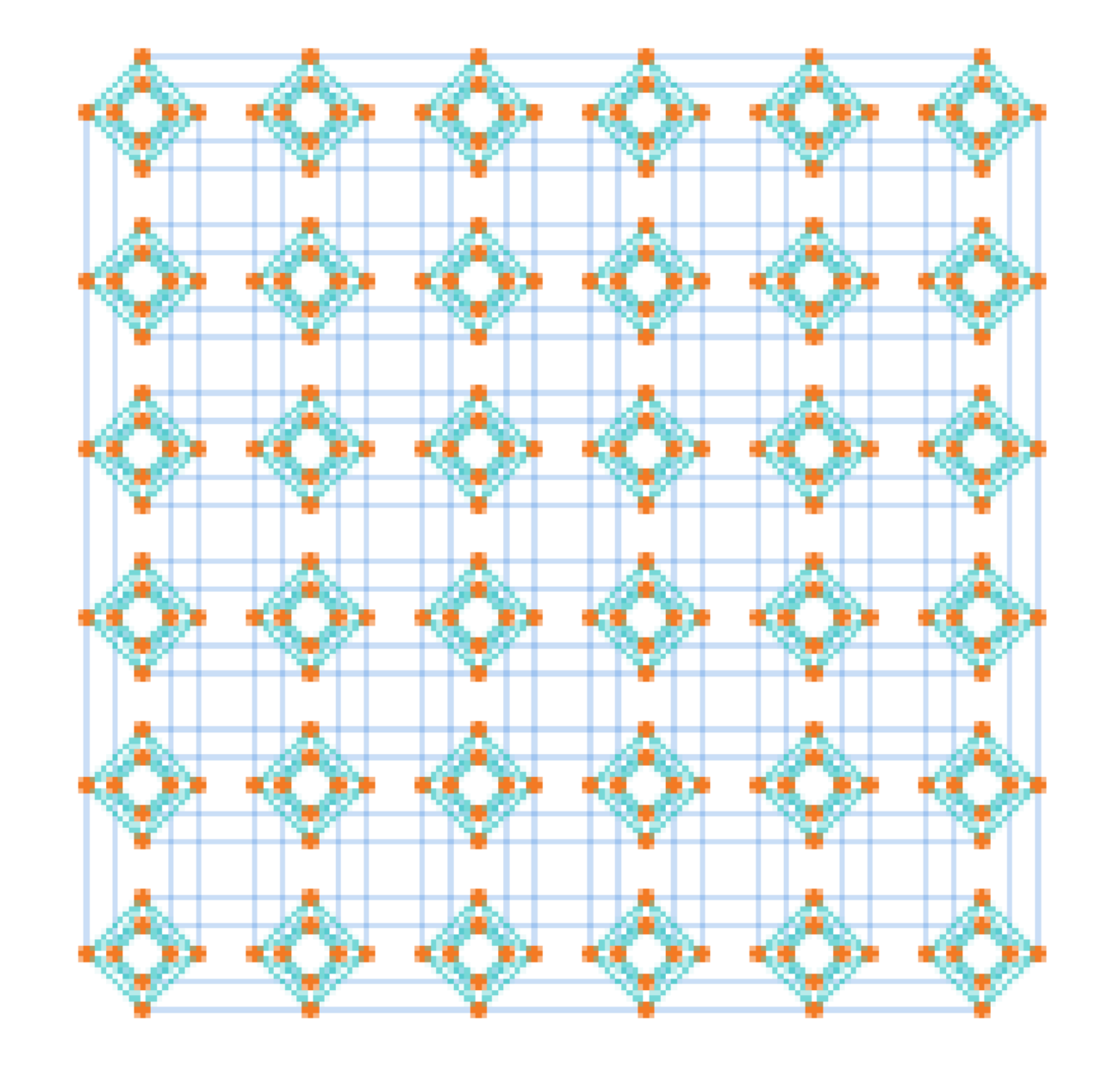}
	\caption{A C6 Chimera graph with 36 unit cells containing 288 qubits. The full D-Wave 2000Q QPU consists of over 2000 qubits with this topology. Adapted from the D-Wave technical manual \protect\cite{dwave_doc}. }
	\label{dwavetopology_fig}
\end{figure}

On a physical quantum annealer such as the D-Wave machine\footnote{Run-time on D-Wave machines was via AWS (Amazon Web Services).} used for this work, 
a representation of the system in terms of logical qubits insufficiently
describes the problem.  On a real device, one must embed the logical qubits 
onto physical qubits that carry out the computation.  The connectivity of
physical qubits is limited by the topology of the machine architecture \cite{nextGenDwaveTop}.  Qubits on the D-Wave 2000Q QPU are laid out in the C6
Chimera graph — a 6-by-6 grid of unit cells — as shown in Fig. \ref{dwavetopology_fig}. 
The Chimera graph has 6 couplers per qubit.  Couplers link qubits on the device and lead to the quadratic term $Q_{ij}$ in Eqn. (\ref{isingHam}).

In Eqn. (\ref{isingHam}), $Q_{ij}$ can describe arbitrary connections across logical qubits.  To accommodate this, when we implement the logical QUBO as a problem on a device, we must specify an embedding that corresponds to a many-to-one mapping of physical qubits to each logical qubit.  Thus chains of many physical qubits are linked together to act as one logical qubit described by each Pauli spin matrix in Eqns. (\ref{dwaveinit}-\ref{isingHam}).   If the chain strength is not large enough, the physical qubits in the chain will not take the same value after annealing, leading to ``broken'' chains.  Thus the chain strength should be high enough to prevent many broken chains.  However, weights in the QUBO are auto-scaled to a range of $[-1,1]$ before the problem is embedded on the physical device.  Chain strengths that are too large will lead to the parameters of the optimization problem of interest to be too small and cause a degradation in the quality of solution \cite{dwave_chain_strength}. The optimal embedding for a given problem, as well as the choice of chain strength, is still a subject of active research. In the section that follows, we discuss  
the effect of system size and chain strength on the quality of the
annealed solution to the $k$-rooks problem. 

%%%%%%%%%%%%%%%%%%%%%%%%%%%%%%%%%%%%%%%%%%%%%%%%%%%%%
\section{Results --- the $k$-rooks problem}\label{results}
%%%%%%%%%%%%%%%%%%%%%%%%%%%%%%%%%%%%%%%%%%%%%%%%%%%%

The  MTDA problem with no false alarms and no missed detections can be modeled as a $k$-rooks problem \cite{k-rooks}. 
The ground state of the corresponding $k$-rooks Ising Hamiltonian has a ground state 
degeneracy of $k!$ -- corresponding to the $k!$ valid solutions that satisfy the constraints 
imposed by Eqns. (\ref{krooksrbaaa}-\ref{krookLinearTerm}) below.

We wish to represent the Ising form $H_P$ of the $k$-rooks problem in order to solve it using AQC. We first define the following matrices:
\begin{align}
  {\mathbf{1}}(k) &\equiv k \times 1  
     \text{ col vector of ones} \nonumber  \\
          &=  \begin{pmatrix} 1 & 1 & 1 & \cdots & 1 \end{pmatrix}^T    \label{oneKcolVecrba}  \\
   I(k) &\equiv k \times k \text{ identity matrix}    \label{identityKrba} \\
     %  &=  \begin{pmatrix} 1 & 0 & 0 & \cdots & 0 \\ 
     % 0 & 1 & 0 & \cdots & 0 \\
     % 0 & 0 & 1 & \cdots & 0 \\ 
     % \vdots & \vdots & \vdots & \vdots & \vdots \\ 
     %  0 & 0 & 0 & \cdots & 1 \\  
     % \end{pmatrix}   \\
   J(k) &\equiv {\mathbf{1}}(k) {\mathbf{1}}(k)^T - I(k)  \nonumber  \\ 
   &=  k \times k \text{ matrix of ones, with zero diagonal}   \nonumber  \\
  &=   \begin{pmatrix} 0 & 1 & 1 & \cdots & 1 \\ 
   1 & 0 & 1 & \cdots & 1 \\
      1 & 1 & 0 & \cdots & 1 \\ 
      \vdots & \vdots & \vdots & \ddots & \vdots \\ 
            1 & 1 & 1 & \cdots & 0 \\  
            \end{pmatrix}   
\end{align}

Using notation similar to that of \cite{k-rooks},
the quadratic term in the final Ising Hamiltonian in Eqn. (\protect\ref{isingHam}) for the $k$-rooks problem is given by
\begin{equation}  \label{krooksrbaaa}
    \qqkrooks = \WrFromKrooks   +   \WcFromKrooks  \,,   %{{\mathbf{Q}}} = -\frac{1}{2} \left( \mathbf{W}_r + \mathbf{W}_c \right),
\end{equation}
and the linear term is given by
\begin{equation}     \label{krookLinearTerm}   
\qkrooks =  \thetaRKrooks + \thetaCKrooks  \, .   
\end{equation}
The $k^2 \times k^2$ matrix 
\begin{equation}     \label{qweqweA}
\WrFromKrooks \equiv I(k) \otimes J(k)
\end{equation} 
and the 
$k \times 1$ column vector 
\begin{equation}  \label{qweqweB}
\thetaRKrooks \equiv (2k-4){\mathbf{1}}(k^2)
\end{equation} 
constrain the rows to have one rook each.
Similarly, 
\begin{equation}   \label{qweqweC}
\WcFromKrooks \equiv J(k) \otimes I(k)
\end{equation} 
and 
\begin{equation}   \label{qweqweD}
\thetaCKrooks \equiv (2k-4){\mathbf{1}}(k^2)
\end{equation} 
constrain the columns to have one rook each.

  %and
  %\begin{equation*}
  %  \qmtdaVec_{ij} = 
  %  \begin{cases}
  %    2M + 2N - 4, & \text{if}\ i>0, j>0 \\  2N - 2, & \text{if}\ i=0, j>0 \\ 2M - 2, & \text{if}\ i>0, j=0   \\  2, & \text{if}\ i=0, j=0   \,.
  %  \end{cases}
  % \end{equation*}
%%%%%%%%We set $\qmtdaVec_{00} = 2$ somewhat arbitrarily in order to push the final solution towards setting the to-be-discarded-anyway $S_{00}=-1$, which reduces ground state degeneracy in BLAH BLAH.

\begin{figure}[t]
    \centering
	\includegraphics
	[width=0.48\textwidth]
	{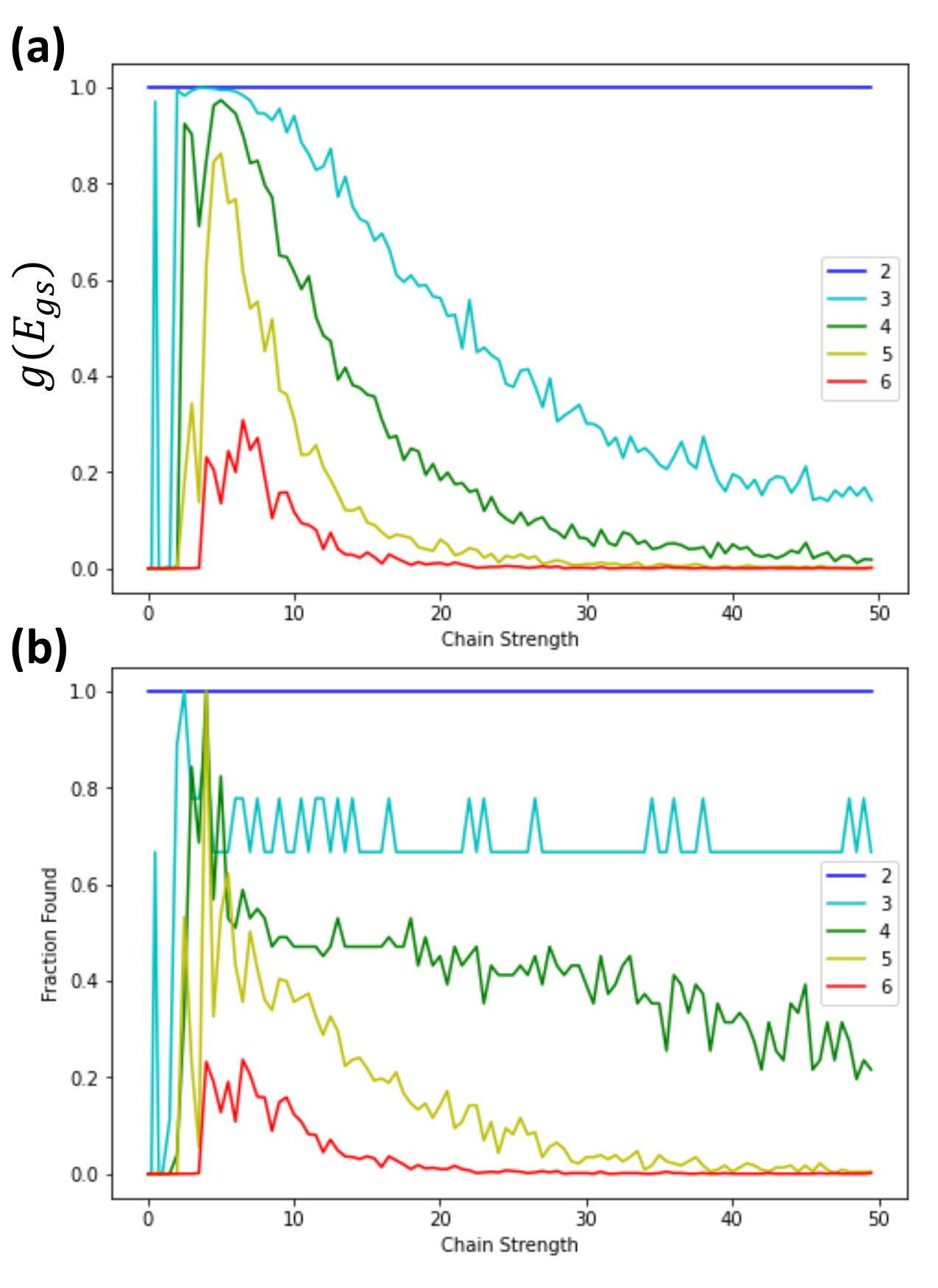}
	\caption {Ground state annealing of $k$-rooks using a D-Wave 2000Q QPU. {\textbf{(a)}} Fraction of shots $g(E_{gs})$ that return the ground state (with energy $E_{gs}$) as a function of chain strength for $k = 2,...,6$ for $N_s = 10^3$ shots per chain strength. {\textbf{(b)}} Fraction of the $k!$ degenerate ground states that are reached by annealing $N_s = 10^3$ shots as a function of chain strength for $k = 2,\ldots,6$.  }
	\label{krook_chainstrength_fig}
\end{figure}

By exact numerical diagonalization \cite{sandvik}, it is straightforward to show that the energy gap between all manifolds of eigenstates of the $k$-rooks Hamiltonian is constant and equal to $8$ in the dimensionless units of Eqn. (\ref{isingHam}).  Since violations of the adiabatic condition in Eqn. (\ref{adiabThm}) depend on the annealing rate and the energy gap, by fixing these we may study the effects of system size, chain strength, and embedding.  The $k=2$ rooks problem comprises $4$ logical qubits and embeds completely on one unit cell without requiring chaining.  Consequently, the D-Wave 2000Q annealer robustly finds the ground states in all of over $10^5$ shots spread across multiple runs performed in this work.  

As we increase the system size, the fraction of shots $g(E_{gs})$ that return the ground state (with energy $E_{gs}$) decreases from unity. $g(E_{gs})$ also begins to depend more sensitively on the chain strength of physical qubits.  In Fig. \ref{krook_chainstrength_fig}a, we show $g(E_{gs})$ as a function of chain strength for $k = 2,...,6$ and for $N_s = 10^3$ shots per chain strength.  We see that for $k > 2$, fewer  shots are annealed into the ground state.  We see that consistently a chain strength of approximately $5$ maximizes $g(E_{gs})$.  Due to the combinatorial growth of the ground state degeneracy, increasing $g(E_{gs})$ increases the probability of finding a larger fraction of each of the $k!$ ground states, as shown in Fig. \ref{krook_chainstrength_fig}b.

In general, for highly degenerate ground states, it is challenging to obtain the entire ground state manifold, even by annealing a large number of shots using a long annealing time.  However, the relative success of AQC is remarkable in comparison with a classical brute force search over the space of all possible solutions.  For the 6-rooks problem, there are $6! = 720$ ground states and over $6 \times 10^{10}$ possible solutions.  In Fig. \ref{krook_chainstrength_fig}b, we see that $N_s = 10^3$ shots find over 20\% of the ground state manifold, 
orders of magnitude more than we would expect by chance.  
For fixed annealing parameters (such as the annealing time $t_f$), increasing $n_s$ returns a fraction of the total number 
of degenerate ground states that approaches unity.

\begin{figure*}[t]
    \centering
	\includegraphics
	[width=1.\textwidth]
	{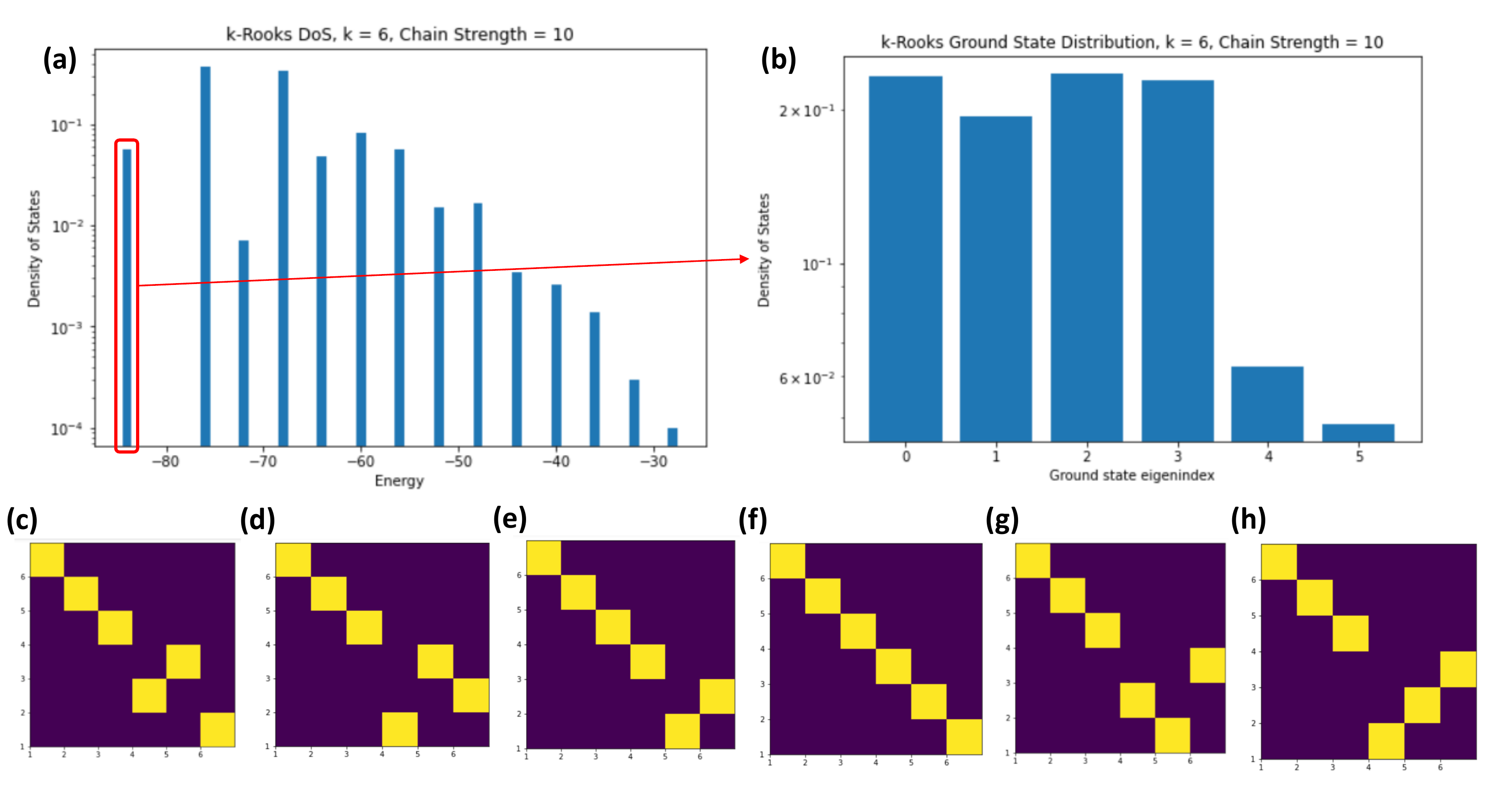}
	\caption {{\textbf{(a)}} Annealed density of states for the biased 6-rooks problem in Eqn. (\ref{krookbiasham}) with $\gamma_0 = 6$ and $m = 3$.  Annealing done on a D-Wave 2000Q QPU using a 50 $\mu$s annealing time, chain strength of 10, $N_s = 10^4$ and composite embedding. {\textbf{(b)}} Distribution of the $ (k-m)! = 6$ states in the ground state manifold.  {\textbf{(c-h)}} Individual ground states corresponding to states 0 through 5 in (b). Yellow squares correspond to rook placements.}
	\label{krook_bias_fig}
\end{figure*}

By adding a bias of the form
\begin{equation}
\label{krookbiasham}
    H_{bias} = -|\gamma_0| \sum_{i=1}^{m} \sigma_{3}^{i(k+1)}.
\end{equation}
to the $k$-rook final Hamiltonian given by Eqn. (\ref{isingHam}), we effectively 
add a negative on-site energy $-|\gamma_0|$ 
to the diagonal of the first $m$ rows and columns. 
This helps enforce a final ground state solution that places a rook in the 
diagonal row-column positions $(1,1), \ldots, (m,m)$, reducing the dimension of the $k$-rooks ground state from $k!$ to $(k-m)!$.

In Fig. \ref{krook_bias_fig}, we show the annealed solution for the biased $k$-rooks problem for case of $k = 6$, $\gamma_0 = 6$, $m = 3$.  To obtain the annealed solution, we use a D-Wave 2000Q QPU with a composite auto-embedder to map the logical qubits onto physical qubits in the device.  The dimension of the Hilbert space of the $k = 6$ rook problem is $2^{k^2} \approx 6 \times 10^9$ while the ground state for $k = 6$, $m = 3$ has degeneracy of $(k-m)! = 6$.  We find all ground states (Fig \ref{krook_bias_fig}b-h) using a single run of $N_s = 10^4$ shots.
%{\color{blue} (a) Is $2^{k^2}$ the \emph{size} or %\emph{dimension} or \emph{whatever} of the Hilbert space? (b) Is it even a Hilbert space?}
%{\color{red} (a) fixed (b) yes}

\section{Results --- General MTDA Problem} \label{MTDAresults}

\begin{figure*}[th]
    \centering
	\includegraphics
	[width=1.\textwidth]
	{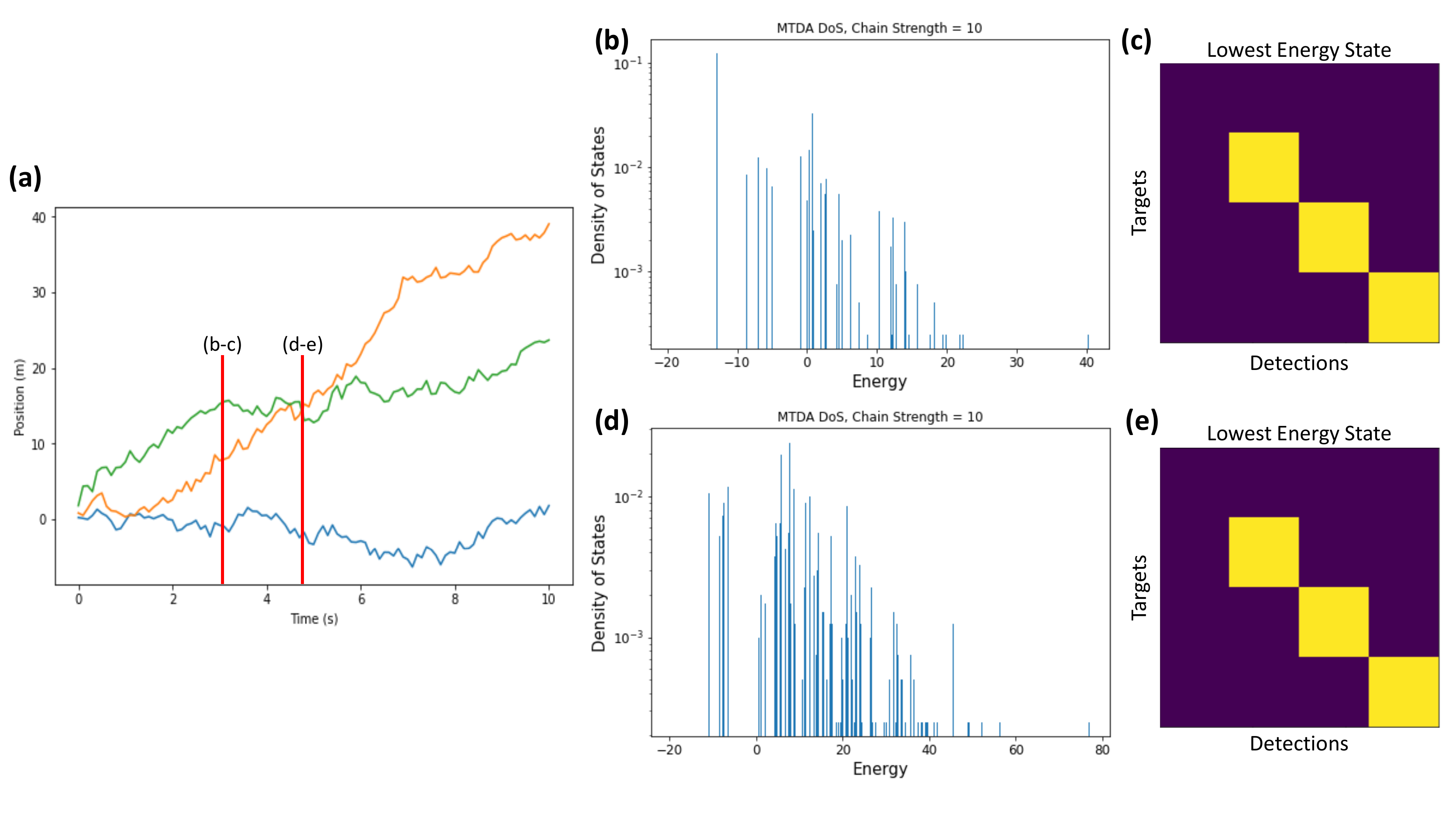}
	\caption {{\textbf{(a)}} Scenario with a linear Gaussian motion model. {\textbf{(b-e)}} MTDA problem with 3 targets and 3 detection annealed on a D-Wave 2000Q QPU using a 200 $\mu$s annealing time, chain strength of 10, $N_s = 5000$ and composite embedding. {\textbf{(b)}} Density of annealed states for scenario time $t = 3 \textrm{s}$.  {\textbf{(c)}} Lowest energy annealed state at $t = 3\textrm{s}$ with correct target and detection association. The first row and column of the state are missed and false alarms respectively. Spin up states (shown in yellow) correspond to estimated associations for the blue, orange, and green (rows 2 through 4 respectively) targets in (a). {\textbf{(d)}} Density of annealed states for scenario time $t = 4.4 \textrm{s}$. 
	{\textbf{(e)}} Lowest energy annealed state at $t = 4.4\textrm{s}$ with same rows and columns as in (c).  We see that when measurements are close together, the QPU annealed solution still allows for disambiguation for large enough cost constraint values $c$.  }
	\label{mtda_fig_3tgt_3det_fig}
\end{figure*}

The adiabatic programming formulation of the MTDA problem leads to two components in the final Ising Hamiltonian in Eqn. (\ref{isingHam}): 
\begin{enumerate}[leftmargin=*,align=left]
  \item A quadratic term that constrains the association of at most one target to each detection and at most one detection to each target.
  \item A linear term that accounts for the cost of target-detection mis-assignment based on the negative log-likelihood.  
      This term also contains the linear part of the constraints mentioned above, similar to Eqn. (\ref{krookLinearTerm}).
\end{enumerate}
Let $\gamma = \text{vec}(\costQ)$ be as defined by Eqn. (\ref{modifiedeqnforty}), 
and let $c , \tilde{c} > 0$ be given.
Then the first term above leads to a modified form of the $k$-rooks cost matrix
\begin{equation} \label{mtda_cost}
 \qqmtda = c \, (\WrFromMTDA + \WcFromMTDA), 
\end{equation}
and the second term gives
\begin{equation}     \label{mtda_cost_linear}
    \qmtda = \tilde{c} \, (\thetaRMTDA + \thetaCMTDA) + \gamma  \, ,   
\end{equation}
where $\WrFromMTDA$, $\WcFromMTDA$, $\thetaRMTDA$, and $\thetaCMTDA$
(defined below) are similar 
to the corresponding $k$-rooks constraints 
but are modified to allow for missed detections and false alarms. 
The cost constraint coefficients $c$ and $\tilde{c}$ 
set the overall energy scale of the constraint terms;
higher values correspond to more aggressive enforcement of the constraints.

With a minor tweak to Eqns. (\ref{oneKcolVecrba})  and (\ref{identityKrba}), 
we define the following matrices: 
\begin{align}
   {{\mathbf{1}}_0}(k) &\equiv k \times 1  
     \text{ col vector of ones with zero in first entry} \nonumber  \\
          &=  \begin{pmatrix} 0 & 1 & 1 & \cdots & 1 \end{pmatrix}^T    \label{oneKcolVecrbaZero}    \\         
    I_0(k) &\equiv k \times k \text{ identity matrix with zero in } (1,1) \text{ entry}  \nonumber  \\
  &=  \begin{pmatrix}   
  0 & 0 & 0 & \cdots & 0 \\ 
   0 & 1 & 0 & \cdots & 0 \\
      0 & 0 & 1 & \cdots & 0 \\ 
      \vdots & \vdots & \vdots & \vdots & \vdots \\ 
            0 & 0 & 0 & \cdots & 1 \\  
            \end{pmatrix}             \label{identityKrbaZero}
\end{align}
%%Compare Eqns. (\ref{oneKcolVecrbaZero})  and (\ref{identityKrbaZero}) with Eqns. (\ref{oneKcolVecrba})  and (\ref{identityKrba}), respectively.

For $N \geq 1$ targets and $M \geq 0$ measurements,
\begin{align}
\WrFromMTDA &= I_0(M+1) \otimes J(N+1)   \nonumber  \\
\WcFromMTDA &= J(M+1) \otimes I_0(N+1)   \nonumber  \\
\thetaRMTDA &= \left(2N-2\right)\Big[{{\mathbf{1}}_0}(M+1) \otimes {\mathbf{1}}(N+1)\Big]    \nonumber \\
\thetaRMTDA &= \left(2M-2\right)\Big[{\mathbf{1}}(M+1) \otimes {{\mathbf{1}}_0}(N+1)\Big]     \,.    \label{mtdarbaaa}
\end{align}
We note that the MTDA constraint matrices 
in   % as given by
Eqn. (\ref{mtdarbaaa}) provide the correct expression of the matrices given in \cite{mtda}. 
%fix their Wc and Wr matrices and add linear terms that they did not have  .  be matter of fact. Give the correct forms and
%note that this corrects the expression given in [1]. 

It is useful to define the
$(N+1) \times (M+1)$ matrix $\qmtdaVec = \left(\qmtdaVecEntry_{ij}\right)$,  %For $(i,j) \neq (0,0)$, 
\begin{equation}
\qmtdaVecEntry_{ij} = (2N-2){\mathbbm{1}}_{j>0}(j) +(2M-2){\mathbbm{1}}_{i>0}(i)  ,
\end{equation}
where the indicator function ${\mathbbm{1}}_{i>0}(i)$ equals one if $i>0$ and equals zero otherwise 
(similarly for $j$).  %${\mathbbm{1}}_{j>0}(j)$ is defined similarly.
It turns out that the linear portion
of the MTDA constraints can be expressed
\begin{equation}
    \thetaRMTDA + \thetaCMTDA = \text{vec}(\qmtdaVec) .
\end{equation}
%where the vectorization $\text{vec}(\qmtdaVec)$ of $\qmtdaVec$ is defined to be
%the $(N+1)(M+1) \times 1$ column vector obtained by concatenating the columns of $\qmtdaVec$.

Incidentally, letting $M=N=k-1$ in the MTDA problem
of Eqns. (\ref{mtdarbaaa}), 
% and noting that $\qkrooks$ in Eqn. (\ref{krooksrbaaa}) can be written
% \begin{equation*}
% \qqkrooks =   (2k-4) \Big[{\mathbf{1}}(k) \otimes {\mathbf{1}}(k)
%      +   {\mathbf{1}}(k) \otimes {\mathbf{1}}(k) \Big] ,
% \end{equation*}
the similarity with the $k$-rooks problem of Eqns. (\ref{qweqweA}-\ref{qweqweD}) is apparent:
\begin{align}
\WrFromMTDA &= I_0(k) \otimes J(k)  \nonumber  \\
\WcFromMTDA &= J(k) \otimes I_0(k)    \nonumber  \\
\thetaRMTDA &= \left(2k-4\right)\Big[{{\mathbf{1}}_0}(k) \otimes {\mathbf{1}}(k)\Big]    \nonumber \\
\thetaRMTDA &= \left(2k-4\right)\Big[{\mathbf{1}}(k)\otimes {{\mathbf{1}}_0(k)} \Big]      \,.    \nonumber
\end{align}

%{\color{red} ANYONE?? Mention how we had to fix their Wc and Wr matrices and add linear terms that they did not have (in a nice way)? Yes, but be matter of fact.  Give the correct forms and note that this corrects the expression given in [1]. }

The terms defined above, in addition to the chain strength that couples physical qubits, set the energy scales of the problem and therefore determine the quantum dynamics on the annealed Ising Hamiltonian.  

Unlike the $k$-rooks system, where our analysis was simplified by the high degree of symmetry of the problem, the linear term in the Hamiltonian depends on the noisy measurements and filtered state estimates, and it  generally leads to an Ising Hamiltonian with on-site disorder \cite{random_ising}.

In Fig. \ref{mtda_fig_3tgt_3det_fig}a, the position as a function of time is plotted 
for an example scenario of 3 targets 
with a linear Gaussian motion model.  We take a state space comprising a position and
velocity in one dimension and we assume  the position is measured 
at regular time intervals according to a linear Gaussian measurement model. 
We set the measurement variance 
$\sigma_M^2 = 0.1 \textrm{m}^2$, 
the probability of detection $p_d = 0.95$,
the clutter Poisson mean $\lambda = 1$, 
and the measurement space volume $\fov = \lengthmeasspaceY$;
see  Eqn. (\ref{modifiedeqnforty}).
Furthermore, the cost constraint coefficients are set to $c = 10$ and $\tilde{c} = 1$.  
The initial positions and velocity of the $i$-th target are given by
\begin{equation}
    \label{initKinematics}
    \left(x_0^{(i)}, \dot{x}_0^{(i)}\right) = \left((i - 1)\ \textrm{m}  \,,\, 2(i - 1)\ \frac{\textrm{m}}{\textrm{s}}\right) \,.
    %\begin{pmatrix}
     %(i - 1)\ \textrm{m}, & 2(i - 1)\ \frac{\textrm{m}}{\textrm{s}}
   % \end{pmatrix} 
\end{equation}
Unless stated otherwise, all further numerical 
experiments in this work were carried out with these parameters.

%{\color{red}TIM-- technically speaking, to fully describe the motion in Fig. \ref{mtda_fig_3tgt_3det_fig}a,
%we still need two more pieces of info for EACH of the three targets. (1) First, we need the process noise covariance matrices.
%If these $\kalmanQ$ matrices are equal for all three targets, then filling in the blue part in Sect. 2 will 
%take care of that. (2) Second, we need the initial state $(pos, vel)$ at reference time $t_0 = 0$ for each target.}

We estimate the target-detection association for three detections on these targets at $t = 3\textrm{s}$ and $t = 4.4\textrm{s}$ by annealing a system prepared in the ground state of Eqn. (\ref{dwaveinit}) to the final MTDA Ising Hamiltonian using a D-Wave 2000Q QPU.
We see the effects of disorder in the energy spectra: the highly degenerate energy eigenstates of the $k$-rooks Ising Hamiltonian are broadened into the bands of states observed  in Fig. \ref{mtda_fig_3tgt_3det_fig}b and d.

At time $t = 3\textrm{s}$, the targets are relatively well-separated compared with the measurement covariance and the lowest energy annealed state (left-most in Fig. \ref{mtda_fig_3tgt_3det_fig}b) that we observe. Fig. \ref{mtda_fig_3tgt_3det_fig}c correctly associates each detection with each target.  The first row and column of the state are missed and false alarms respectively. Spin up states (shown in yellow) correspond to estimated associations for the blue, orange, and green (rows 2 through 4 respectively) targets in Fig. \ref{mtda_fig_3tgt_3det_fig}a.  

In Fig. \ref{mtda_fig_3tgt_3det_fig}d and e, we consider the MTDA annealed solution at the later time $t = 4.4 \textrm{s}$ for the same scenario.  As the distance between target 3 (shown in Fig. \ref{mtda_fig_3tgt_3det_fig}a in green) and the other targets becomes large relative to the measurement covariance and accounts for the overall shift of the annealed density of states to higher energy seen in Fig. \ref{mtda_fig_3tgt_3det_fig}d.  We also find smaller support for the lowest energy state.  
In a classical ILP problem, each target can be associated with at most one target and conversely.  In the quantum annealing problem, 
this constraint is enforced energetically by 
% Eqn. (\ref{mtda_cost}) and the overall scale of this constraint is set by the parameter $c$.  
Eqns. (\ref{mtda_cost})-(\ref{mtda_cost_linear}) and the overall scale of this constraint is set by the parameters $c$ and $\tilde{c}$. 
In the above scenario, we have set $c = 10,\ \tilde{c} = 1 $, 
which are large enough values 
(relative to the cost matrix in Eqn. (\ref{modifiedeqnforty})) 
to energetically enforce target/detection disambiguation.

\subsection{Larger MTDA Systems and False Alarms}

\begin{figure}[t]
    \centering
	\includegraphics
	[width=0.5\textwidth]
	{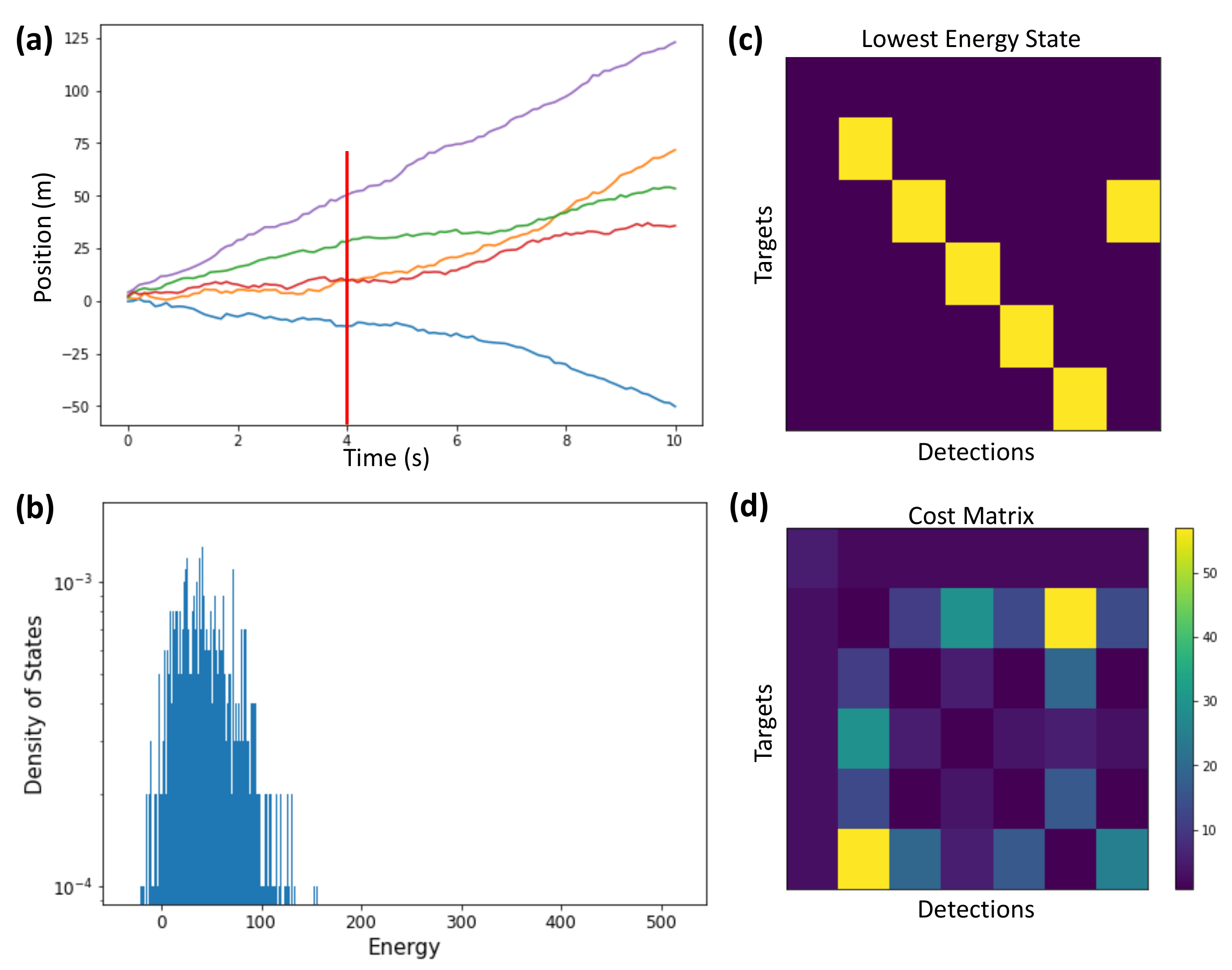}
	\caption { {\textbf{(a)}} Scenario with a linear Gaussian motion model. MTDA problem with 5 targets and 6 detections annealed on a D-Wave 2000Q QPU using a 100 $\mu$s annealing time, chain strength of 10, $N_s = 10^4$ and composite embedding.  The 6th detection is a false alarm located near the crossing of targets 2 and 3.  {\textbf{(b)}} Density of annealed states for scenario time $t = 4 \textrm{s}$. {\textbf{(c)}} Lowest energy annealed state at $t = 4 \textrm{s}$. {\textbf{(d)}} Cost matrix in Eqn. (\ref{modifiedeqnforty}) for scenario and time. }
	\label{mtda_fig_5tgt_6det_fig}
\end{figure}

Since the target and detection assignment constraints are enforced energetically at the level of the Hamiltonian, the quantum annealed solution allows us to capture the ambiguity in scenarios where uncertainty is high.  In Fig. \ref{mtda_fig_5tgt_6det_fig}a, we show a scenario with 5 targets and 6 detections -- one detection per target and a single false alarm near the crossing of targets 2 and 3. 
%{\color{red}Is this the only false alarm (at the time of crossing)? Prob fine, but I got a little confused.}
Fig. \ref{mtda_fig_5tgt_6det_fig}b shows the annealed density of states found using $N_s = 10^4$ shots on a D-Wave 2000Q QPU using a 100 $\mu$s annealing time.  We note that as the system grows relative to the 3 target, 3 detection scenario shown above, the typical gap between energy levels shrinks.  This leads to a smaller probability of annealing into the actual ground state of the final Ising Hamiltonian.

Fig. \ref{mtda_fig_5tgt_6det_fig}c shows how the annealed solution at $t = 4 \textrm{s}$ for the same scenario captures the ambiguity by assigning both measurements (detections 2 and 6) to target 2.  In the traditional ILP solution of the MTDA problem, this would not be possible. Instead, the constraints would enforce the solution to pick one of the possibilities -- either detection being assigned to target 2.  In the QPU annealed solution, the energetic constraints on the solution space cause the system to favor at most one detection per target (and vice versa), but this competes with the cost matrix in Eqn. (\ref{modifiedeqnforty}) and shown in Fig. \ref{mtda_fig_5tgt_6det_fig}d.  For low false alarm rates 
(in this scenario $\lambda = 1$ and $\fov = \lengthmeasspaceY$), 
the cost matrix leads to an on-site potential from the linear term in Eqn. (\ref{isingHam}) that may be higher than the 
energy penalty 
%from the   % quadratic (RBA removed "quadratic" since there is a linear portion of the constraints)
% from the constraints from assigning two BLAH BLAH
for assigning two detections to the same target.  
Increasing the value of $c$ strengthens this constraint, but since the D-Wave annealer scales all energy parameters such that they lie between $[-1,1]$, too high values of $c$ then dominate the behavior of the lowest energy annealed state. This leads to a degeneracy of ground states directly analogous to the $k$-rooks solutions in the large $c \rightarrow \infty,\ \tilde{c} \rightarrow \infty$ limit. 
%{\color{red}(TALK ABOUT $\tilde{c}$ HERE AS WELL?)}

\section{Numerical Study of Anneal Time}\label{numerical}

The anneal time is a central parameter of adiabatic quantum computing.  Numerical computations of the adiabatic evolution of a small example MTDA system performed on a classical computer show that by increasing the anneal time, the probability of occupying the ground state approaches unity. In practice, when analyzing output of our calculations on D-Wave, we find that dependence of the minimal annealed state energy on the annealing time is more complicated and depends on the system size.

\subsection{Solution on a Digital Computer}

\begin{figure*}[t]
    \centering
	\includegraphics
	[width=1\textwidth]
	{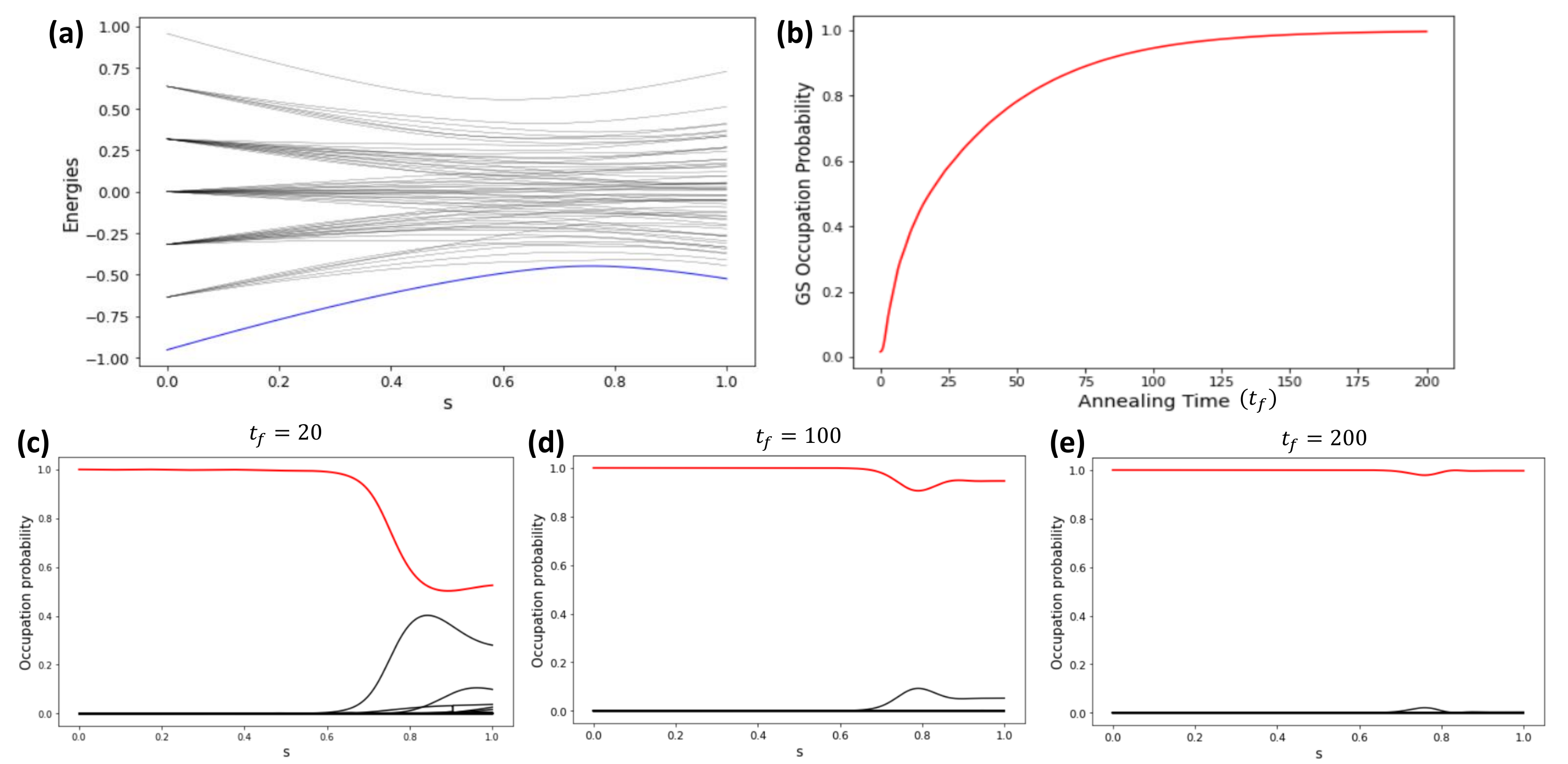}
	\caption {\textbf{(a)} Energy eigenvalues of the instantaneous MTDA Hamiltonian as calculated using QuTiP  \protect\cite{qutip1,qutip2}. Eigenvalues shown as we vary the parameter $s$ from Eqn. (\ref{adiabham}). The scenario consists of one target, located at the origin, and two nearby measurements. The blue curve corresponds to the instantaneous ground state energy.  A gap is present for all values of $s$ indicating validity of the adiabatic theorem for large enough annealing time $t_f$.  \textbf{(b)} The final $s = 1$ occupation probability of the lowest state for the system in (a) as a function of annealing time $t_f$. \textbf{(c,d,e)} State occupation probability as a function of $s$ for an initial state prepared in the ground state of Eqn. (\ref{dwaveinit}) for several values of the annealing time $t_f$.  The red curve shows the evolution of the occupation probability of the state with the lowest energy.  Black curves show occupation probabilities of higher energy states due to adiabatic transitions during system evolution.}
	\label{adiab_ev_1tgt_2det_fig}
\end{figure*}

For small enough system sizes, we can calculate the spectrum exactly by diagonalizing the Hamiltonian on a classical computer.  We take a simplified scenario with a single target with two nearby measurements.  
Using QuTiP \cite{qutip1,qutip2}, we solve the time evolution of the dynamics using the GKSL master equation \cite{intro2Lindblad,weinbergGKSL,gkslG,gkslL} and obtain the instantaneous energies and eigenstates as we evolve the system from Eqn. (\ref{dwaveinit}) at $s = 0$ to the MTDA Ising Hamiltonian at $s = 1$.  We plot the instantaneous eigenvalues as a function of $s$ in Fig. \ref{adiab_ev_1tgt_2det_fig}a with the ground state shown in blue.  A gap between the ground state and excited states is present for all values of $s$ implying that a large enough annealing time $t_f$ can be chosen such that we adabatically remain in the ground state. 

To validate this, we prepare the system to be initially in the ground state of  Eqn. (\ref{dwaveinit}) and then calculate the occupation probability as a function of $s$ as shown in Fig. \ref{adiab_ev_1tgt_2det_fig}c-e for several values of the annealing time $t_f$.  The red curve shows the evolution of the occupation probability of the state with the lowest energy.  Black curves show occupation probabilities of higher energy states due to diabatic transitions during system evolution.  In Fig. \ref{adiab_ev_1tgt_2det_fig}b, we see that the final $s = 1$ occupation probability of the lowest state for the system monotonically increases as a function of the annealing time $t_f$. The precise annealing schedule that D-Wave uses differs slightly in its exact coefficients of $H_B$ and $H_P$, but shares the same monotonicity as the linear convex coefficients that we use in Eqn. (\ref{adiabham}).

The numerical results above provide an upper bound for the performance of a quantum annealer in the idealized adiabatic limit.  In practice there are many reasons why the performance of a physical device would fail to saturate this limit.  Our calculations in Fig. \ref{adiab_ev_1tgt_2det_fig} are carried out at zero temperature; however thermal fluctuations exist in any physical system and may cause jumps from the ground state to a higher energy state.  Since no real-world computation is executed in perfect isolation, the possibility exists for other forms of interactions with the environment that remain unaccounted for in our modeling.

\subsection{Anneal Time Dependence on D-Wave}

To study the dependence of the D-Wave annealed solutions as a function of annealing time, we use a simplified set of measurement and target scenarios of size $(N,M=N+1)$ for $N = 1,\ldots,8$.  For consistency, we spread the targets out uniformly at $1$m intervals and take measurements with covariance of $\sigma_M = 0.1\textrm{m}^2$.  The measurement not associated with any of the targets is a false alarm placed a fixed distance of 1m away from the last target. Similar to the results presented above in Section \ref{MTDAresults}, 
we use an assignment costs $c=10$ and $\tilde{c}=1$,    %we use an assignment cost $c=10$, 
and a chain strength of $10$.  For each of the 11 anneal times considered, 20 runs of $n_s = 500$ shots were performed.  
Shown in Fig. \ref{MTDA_annealTimeDep_fig}a are the smallest energies $E_{gs}$ over all runs for each annealing time $t_f$. We see no dependence on anneal time for system sizes smaller than or equal to $(N,M) = (6,7)$.  For $(N,M) = (7,8)$ and $(N,M) = (8,9)$, there is a weak dependence on $t_f$ leading to a lower value of $E_{gs}$ found.

The observed relative weak dependence on anneal time of both $E_{gs}$, and of the ground state occupation $g(E_{gs})$, gives some evidence that, if the annealing time $t_f$ is driving adiabatic transitions to higher energy states, then we are in the large-annealing time limit to the right in Fig. \ref{adiab_ev_1tgt_2det_fig}(b).  However, the actual values for $g(E_{gs})$ that we obtain from our AQC computations are typically quite low -- typically no larger than $0.1$ -- and yet are essentially independent of annealing time $t_f$.  This provides support for the claim that other failure modes of the device, beyond adiabatic transitions driven by too low a choice of $t_f$, are responsible for the large fraction of higher energy annealed shots.

\section{Extreme statistics}\label{issues}
For a given number of independent shots $n_s$, and a fixed anneal time $t_f$, 
the estimator of the minimum energy $E_{gs}$
%{\color{red}(a little confusing-- what estimator?  may be fine as is)}
is an extreme statistic that follows a (minimum) Gumbel distribution whose PDF is 
\begin{equation}\label{gumbel}
    p(x;\alpha,\beta)=\frac{1}{\beta} \exp\!{\Big[\frac{x-\alpha}{\beta} - \exp
    \!\Big(\frac{x-\alpha}{\beta}\Big)\!\Big]},\,\,x\in\Bbb{R},
\end{equation}
 where the parameters $\alpha$ and $\beta$ are determined by the distribution from which the shots are drawn \cite{extremestats}. Except for small problems, this distribution is unknown and the parameters must be estimated from the samples.  Each ``run'' of an adiabatic computer with $n_s$ shots produces one sample of $\hat E_{0}$.  Several such runs will produce samples from which estimates $\hat \alpha$ and $\hat\beta$  can be computed.  These estimates are functions of the anneal time $t_f$.  Their  utility in AQC remains to be explored.   

\begin{figure*}[t]
    \centering
	\includegraphics
	[width=1\textwidth]
	{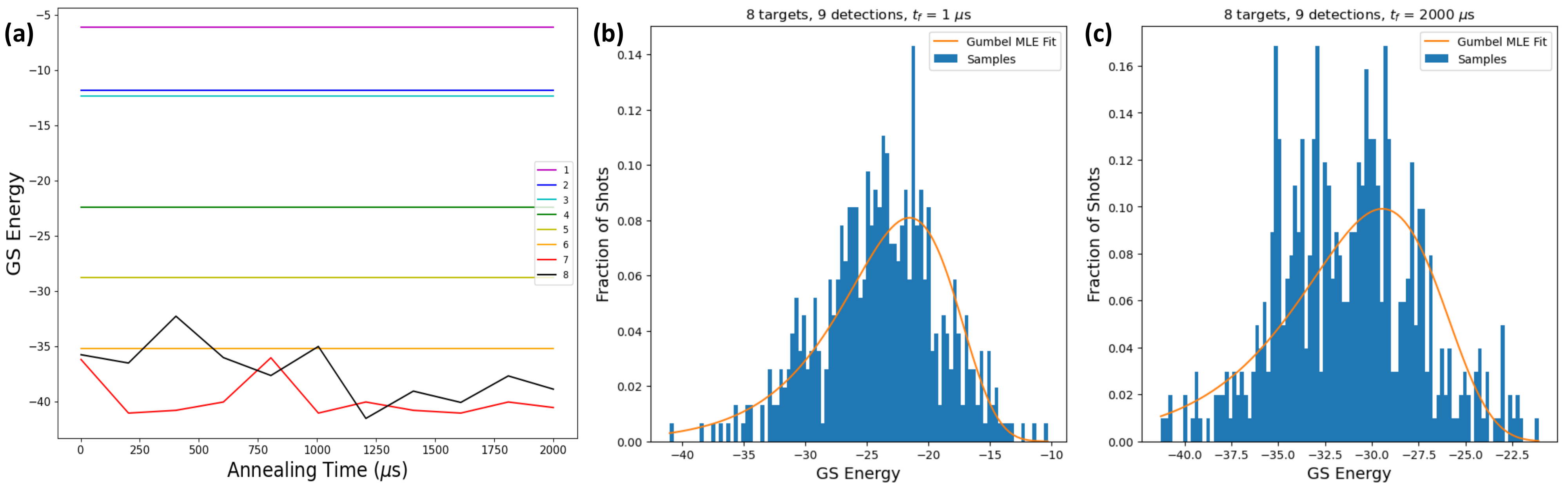}
	\caption {\textbf{(a)} $E_{gs}$ as a function of anneal time $t_f$ for various system sizes (N,M): violet (1,2), blue (2,3), cyan (3,4), green (4,5), yellow (5,6), orange (6,7), red (7,8), and black (8,9).  For a given system size, identical measurements and target location estimates were used.  For each of the 11 anneal times considered, 20 runs of $n_s = 500$ shots were performed.  We see no dependence on anneal time for system sizes smaller than or equal to (6,7).  For (7,8) and (8,9), there is a weak dependence on $t_f$ leading to a lower value of $E_gs$ found. \textbf{(b,c)} normalized histogram of low energy shots with MLE estimate of  Gumbel's  extreme statistic PDF.}
	\label{MTDA_annealTimeDep_fig}
\end{figure*}

In Fig. \ref{MTDA_annealTimeDep_fig}b-c, we show the lowest energy shots per run for 500 runs each with $n_s = 500$ shots 
for the same scenario as used for Fig. \ref{MTDA_annealTimeDep_fig}a, with $(N,M) = (8,9)$.  
The fraction of lowest energy annealed shots are shown in blue for fixed $t_f^{(1)}= 1\ \mu\textrm{s}$ 
(Fig. \ref{MTDA_annealTimeDep_fig}b) and $t_f^{(2)} = 2000\ \mu\textrm{s}$ 
(Fig. \ref{MTDA_annealTimeDep_fig}c).  The maximum likelihood estimate (MLE) best fit of the Gumbel distribution (see Eqn. (\ref{gumbel})) is shown in orange.  For $t_f^{(1)} = 1\ \mu\textrm{s}$, the best fit parameters obtained are 
\begin{equation}
    \label{1musgumbel}
    \hat{\alpha}^{(1)} \approx -21.52,\ \hat{\beta}^{(1)} \approx 4.55,
\end{equation}
while for $t_f^{(2)} = 2000\ \mu\textrm{s}$, we obtain
\begin{equation}
    \label{2000musgumbel}
    \hat{\alpha}^{(2)} \approx -29.43,\ \hat{\beta}^{(2)} \approx 3.71.
\end{equation}

\section{Hybrid Quantum-Classical Computation}\label{hybrid}

In this paper we used an annealing QC to solve the MTDA assignment problem.  This ILP problem is NP-hard and scales poorly on a classical computer as the number of targets and measurements grows.  Using the QC allows us to take advantage of the inherent advantages of quantum processing for doing measure-to-track association.  

The assignment problem is, however, only one step in the Bayesian recursion that forms the basis of most modern tracking systems \cite{barshalomjpda}.  
Completing the recursion to build a practical tracking system involves propagating the target state as defined 
in Eqn. \eqref{eq:jpdaMotUpdate}   
as well as performing a (Bayesian) information update 
via the likelihood function of Eqn. \eqref{eq:jpdaLF}.
These steps of the recursion are appropriate for solution on a classical computer both because of their form and also because they scale linearly in the number of targets once the MTDA assignment problem has been solved.

In this way, the full solution to the tracking problem outlined here would be \textit{a hybrid quantum/classical computational approach}.  Techniques for integrating classical and hybrid computation to solve practical problems will be common as QC enters the mainstream.  This is true of both gate-based and annealing QC.  In the application of QC to solving the MTDA assignment problem, we see one way in which this arises within the context of an annealing QC. A prime example in gated-based QC  is Shor's integer factoring 
algorithm.  It  uses QC to find the ``phase'' of a certain Fourier operator, which is followed by a classical continued fraction expansion algorithm to find the order from the phase, which  in  turn is equivalent to integer factorization. See   \cite[Chap. 5]{mikeandike} for an excellent presentation of the details. 

\subsection{Hybridized Probabilistic Data Association}

When run on an idealized annealer, we can find with arbitrarily high probability the single ground state solution to this MTDA ILP.  Physical devices introduce a variety of noise sources, which, as discussed above, degrade the solution.  
This forces us to sample the annealer to find the desired lowest-energy solution \cite{martinisDwave}.
The question arises: \textit{if we need to collect $10^{3}$ - $10^{4}$ samples to find the ground state--and even then without certainty--what might we do with remaining samples?}

One possibility is to select a subset of these eigenstates--perhaps those with the lowest energies--and compute the posterior likelihood of each according to Eqn. \eqref{ACthreefifteen}.  Normalizing these likelihoods would provide a basis for the probabilistic data association needed to complete a hybridized form of the JPDA recursion.  

Unlike the \textit{hard} assignment decisions generated by the MTDA ILP, the output of this hybridized algorithm would be a probabilistic association between targets and measurements.  This algorithm would be a hybrid in the deeper sense that both the properties of the annealer \textit{and} the classical posterior probabilities inform the updating of the target state.  Understanding the behavior of this type of hybrid algorithm would require an analysis of how the classical posterior probabilities and quantum annealer sampling probabilities relate to one another.

%%%%%%%%%%%%%%%%%%%%%%%%%%%%%%%%%%%%%%%%%%%%%%%%%%%%%
\section{Concluding Remarks}\label{conclsion}
%%%%%%%%%%%%%%%%%%%%%%%%%%%%%%%%%%%%%%%%%%%%%%%%%%%%%
This paper reports the results for using adiabatic QC to solve  single scan MTDA problems.  It validates the theory reported earlier this year in \cite{mtda}.  The results show that the  present state of technology can solve small problems reliably, but that larger problems remain challenging.  Whether or not adiabatic quantum computation provides a speedup \cite{speedup} over other approaches to the MTDA problem remains to be seen.

%{\color{red}The result presented here show ... list what we have learned.}

Other possibilities for applying AQC to multi-target tracking remain to be explored.  Different annealing schedules, including using sudden quenches, have the potential to improve the ground state estimates. Extending the possibilities of modifying the annealing process to improve performance, reverse annealing could be investigated. In reverse annealing, the system is initialized into a classical state and then one modulates the transverse Ising field (given by the initial Hamiltonian $H_B$), in search of a better classical solution to $H_P$ than the initial state \cite{reverseAnneal}.  This could allow for refinement in the optimization around a proposed target/track assignment.

The MTDA problem studied in this paper is for a single scan of data.  Such problems can be solved  exactly using the Hungarian method, which is a class of  combinatorial optimization algorithms that includes  Munkres and  Jonker–Volgenant.  The challenge is that these  methods are not exact for multidimensional assignment problems, that is, for problems in which sequences of feasible measurements over several scans are assigned to targets. Lagrangian relaxation methods work well for these problems but are not guaranteed to be exact.   

As first shown in  \cite{mtda}, AQC has the inherent capability of finding  exact solutions of  multidimensional assignment problems with high probability.  
In addition to the practical challenges that have already been  presented, 
a  fundamental obstacle arises---the number of qubits  needed to solve  multidimensional assignment problems equals the number of feasible measurement sequences.  Thus, Morefield's ILP approach  requires  large QC machine architectures for long sequences and high false alarm rates.  Currently available technology is good enough only for small problems.  Increasing QC   capability will eventually support some practical applications.  Nonetheless, 
sufficiently long sequences and high  enough false alarm rates will overwhelm any available resource.

%%%%%%%%%%%%%%%%%%%%%%%%%%%%%%%%%%%%%%%%%%%%%%%%%%%%%%%%%%%%%%%%%%%%%%%%%%%%%%%%%%%%%%%%%%%%%%%%%
\appendix{. QUBO, Ising and ILP}
\label{QUBOIsingandILP}
% note there is no {} to put a title. Each appendix has its own title
%%%%%%%%%%%%%%%%%%%%%%%%%%%%%%%%%%%%%%%%%%%%%%%%%%%%%%%%%%%%%%%%%%%%%%%%%%%%%%%%%%%%%%%%%%%%%%%%%
%%%%%%%%%%%%%%%%%%%%%%%%%%
Many problems in combinatorial optimization   can be cast in the mathematical form of a quadratic unconstrained binary optimization (QUBO) problem. See \cite{kochenbergerQUBO} for an extensive list.  QUBO problems are defined here as 
\begin{eqnarray}\label{qubo}
& \arg\mathop{\min} \quad x^T Q x \\
&\textrm{such that  } x\in \{0,1\}^n , \nonumber
\end{eqnarray}
where $Q$ is an $n \times n$ symmetric matrix of real coefficients. The binary solution of the QUBO is the minimum energy of an Ising model with $n$ sites, and conversely.  To see this, recall that the Ising model  Hamiltonian is  
\begin{eqnarray}\label{ising}
& H(\sigma)=-\sigma^T J \sigma - \mu h^T \sigma,\quad \sigma\in\{-1,+1\}^n,
\end{eqnarray}
where  $J$ is an $n \times n$ symmetric matrix of real interaction  coefficients, $h$ is a real vector of external magnetic field strengths at the $n$ sites, and $\mu$ is the magnetic moment.  The $(i,j)$ entry of $J$ is zero if the sites are not adjacent.  The diagonal entries of $J$ can be nonzero.   The components of the  vector $\sigma$ are the up/down spins $(+1,-1)$ at the $n$ sites. Let  $\sigma = 2 x - e$, where all the components of $e$ are $1$, so that  $\sigma\in\{-1,+1\}^n$ maps 1-to-1  to $x\in\{0,1\}^n$. Note that  $x_j^2 = x_j$, where $x_j$ is the $j$-th component of $x$,  so that 
\begin{equation}\label{matrixalg}
   \alpha^T x = x^T Diag(\alpha) x 
\end{equation}
for any vector $\alpha$, where $Diag(\alpha)$ is the $n \times n$ diagonal matrix whose $j$th diagonal entry is $\alpha_j$.  Using this identity, it follows that 
\begin{align}
   Q_{\textrm{Ising}}& \equiv H(2x-e)\nonumber\\
   & = x^T\big(-4 J+Diag(4 J e -2 \mu h)\big) x,
\end{align}
where the additive constant $\mu h^T e - e^T J e$ is omitted.  
Thus, $\arg \min$ of  the QUBO with   corresponding 
matrix $Q_{\textrm{Ising}}$ is the minimum energy of the Ising model \eqref{ising}. The converse is shown in the same manner using the inverse mapping $x=(\sigma + e)/2$.

A binary integer linear programming (ILP) problem is any optimization problem that can be cast in the ``standard'' form
\begin{align*}
       \min \quad &c^T x\\
       \textrm{subject to} \quad &A x = b\\
       &x\in\{0,1\}^n,
   \end{align*}      
where $A$ is an $m\times n$ matrix and $b$ is a vector of length $m$, both with integer entries.  It can be posed as a QUBO using  a penalty function to enforce the constraints $A x = b$. To see this, write the matrix equation $A x = b$ as $m$ linear equations $A_j x = b_j, j=1,\ldots,m$, where $A_j$ is row $j$ of $A$. 
The  penalty function is a weighted sum  of the squared constraint violations; using \eqref{matrixalg} and dropping  $\sum_j w_j b_j^2$ gives   
\begin{align*}
    \sum_{j=1}^m w_j &\big(A_j x -b_j\big)^2\\& =\sum_{j=1}^m w_j \,x^T \big(A_j^T A_j -2 Diag(b_j A_j) \big) x, 
\end{align*}
where  $w_j>0$ is the weight for the $j$th equation.  
%{\color{blue}(Where did $x$ go in the RHS of previous Eqn?)}
Adding the penalty   to the objective function  $c^T x = x^T Diag(c) x$  gives the matrix of the QUBO for the binary ILP:   
\begin{align}
    Q_{\textrm{BILP}} = Diag(c) + \sum_{j=1}^m w_j \big(A_j^T A_j -2 Diag(b_j A_j)\big).
\end{align}
The solution of this QUBO converges to the solution of the binary ILP as the weights $w_j$ go (uniformly) to $\infty$.  
Thus, the penalized form of binary ILPs is equivalent to  Ising models and can be solved on  adiabatic quantum computers.

%%%%%%%%%%%%%%%%%%%%%%%%%%%%%%%%%%%%%%%%%%%%%%%%%%%%%%%%%%%%%%%%%%%%%%%%%%%%%%%%%%%%%%%%%%%%%%%%%%%%%%
\acknowledgments
This work was supported by Metron under corporate IRAD funding  as part of a larger Quantum Computing initiative. Other active members of the Metron  group include John Asplund and Ian Herbert. We  thank Tyler Takeshita of AWS  for useful discussions, and AWS for granting us a computational time budget  for this work. We  thank Eric  Ladizinsky and Murray Thom  of D-Wave Systems for help with anneal time questions.  We also thank Brian La Cour of ARL/UT for useful discussions.

%%%%%%%%%%%%%%%%%%%%%%%%%%%%%%%%%%%%%%%%%%%%%%%%%%%%%%%%%%%%%%%%%%%%%%%%%%%%%%%%%%%%%%%%%%%%%%%%%%%%%%
\bibliographystyle{IEEEtran}
%\bibliography{IEEEabr,MyBibFile}

%%%%%%%%%%%%%%%%%%%%%%%%%%%%%%%%%%%%%%%%%%%%%%%%%%%%%%%%%%%%%%%%%%%%%%%%%%%%%%%%%%%%%%%%%%%%%%%%%%%%%%
\thebiography
%% This biostyle allows you to insert your photo size 1in X 1.25in
\begin{biographywithpic}
{Timothy M. McCormick}{mccormickt.png}
received his B.S. in physics from University of Delaware and his Ph.D. in theoretical physics from the Ohio State University, where his research focused on thermoelectric transport and electronic structure of topological semimetals. He is a research scientist at Metron Inc., where his work lies at the intersection of physics, statistics, and computation.  His current interests include underwater acoustics, compressed sensing, multi-agent reinforcement learning, and quantum computation.
\end{biographywithpic} 

\begin{biographywithpic}
{Bryan R. Osborn}{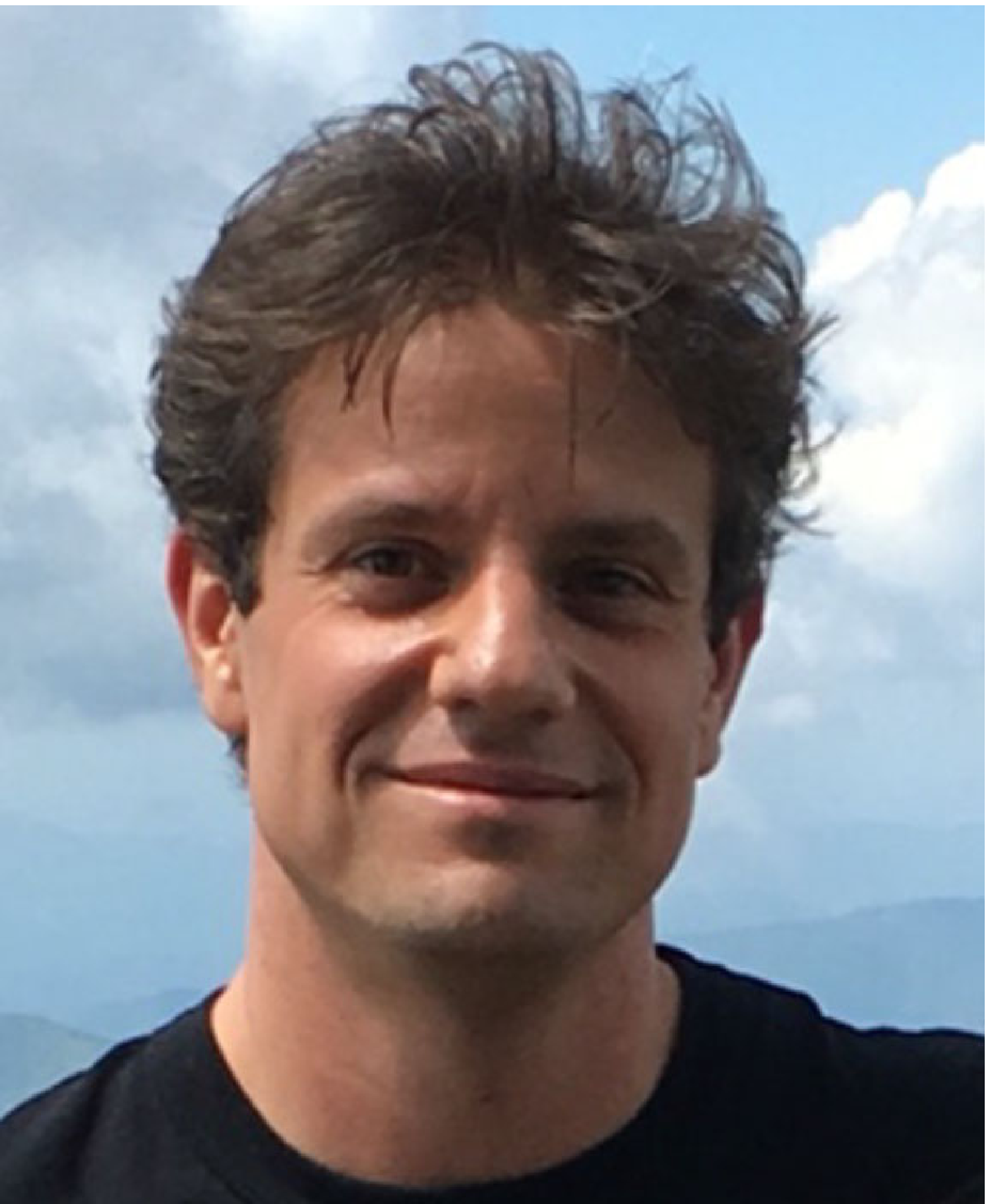}
received  B.S. degrees in Physics and Mathematics in 2001 and an M.S. in Applied Mathematics and Scientific Computation in 2004 from the University of Maryland, College Park.  He is currently a Senior Research Scientist at Metron, Inc. where he leads development and application of tracking algorithms in a variety of contexts.  His interests include distributed sensing systems, high-performance computing, and interactive visualization.
\end{biographywithpic}

\begin{biographywithpic}
 {R. Blair Angle}{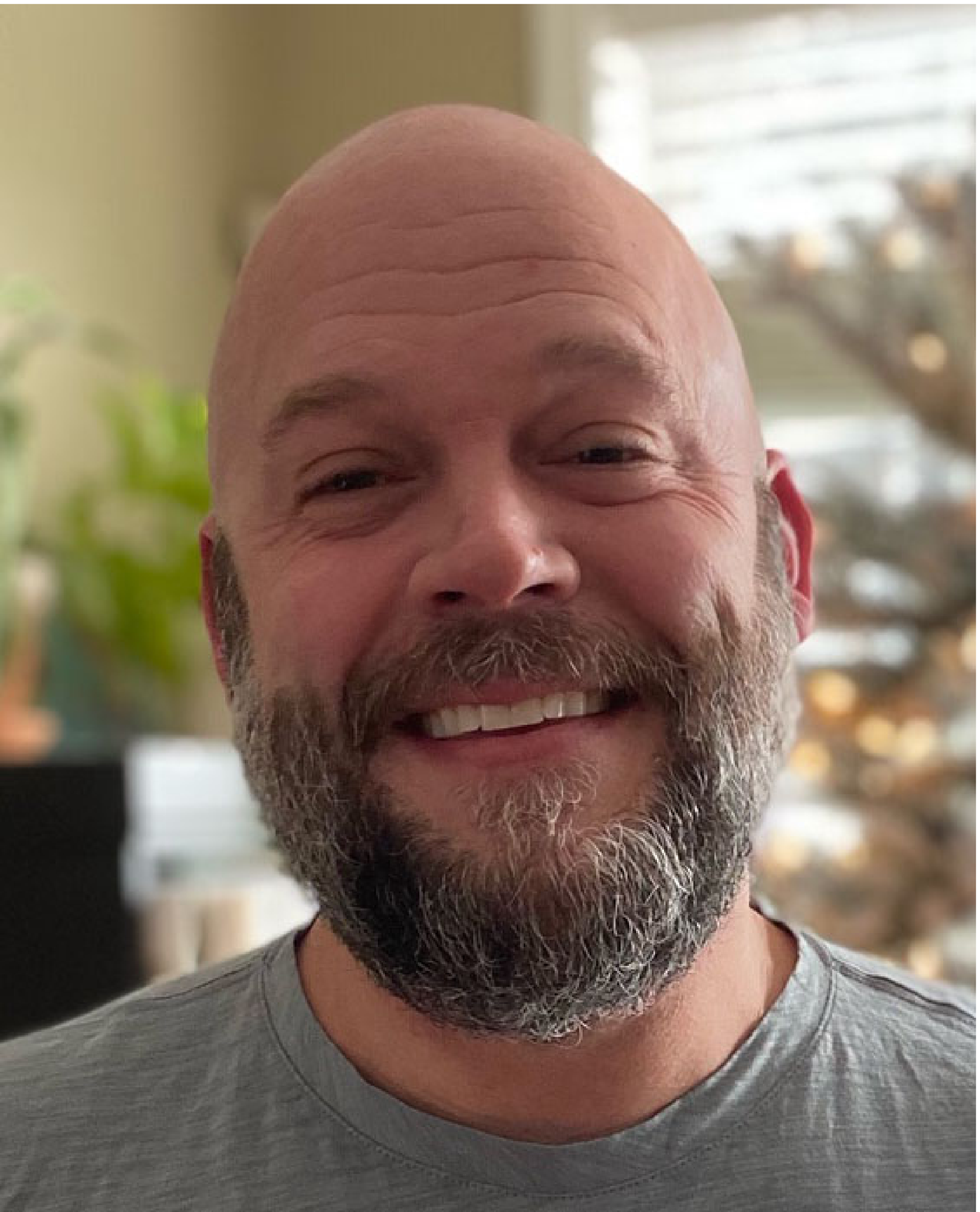}   %{r_blair_angle_photo_2.jpg}    %{blairpic.jpg}
received  his Ph.D. in mathematics from the University of California, San Diego,
and is currently a senior research scientist at Metron, Inc.
His current work involves multi-object tracking, data fusion, analytic combinatorics,
and their interplay.
He recently co-authored the book Analytic Combinatorics for Multiple Object Tracking, Springer, 2021.
\end{biographywithpic}

\begin{biographywithpic}
{Roy L. Streit}{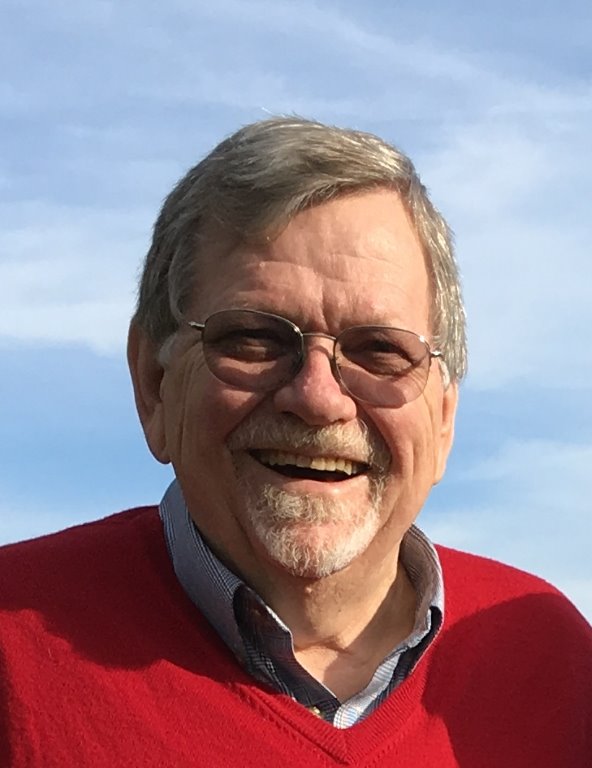}
joined Metron in 2005.  His interests include multi-target tracking, multi-sensor data fusion, signal processing, medical imaging, and business analytics.  His recent work involves applications of analytic combinatorics to multi-target tracking, natural language processing, and subgraph matching in high level fusion.  He is a Life Fellow of the IEEE.  He co-authored Analytic Combinatorics for Multiple Object Tracking, Springer, 2021, and Bayesian Multiple Target Tracking, Second Edition, Artech, 2014.  He is also the author of Poisson Point Processes: Imaging, Tracking, and Sensing, Springer, 2010.  Before 2005, he was in Senior Executive Service at Naval Undersea Warfare Center, Newport, RI.
\end{biographywithpic}

\end{document}